%% file: main.tex
\documentclass[journal,onecolumn]{IEEEtran}
\usepackage{cite}
\usepackage{amsmath,amssymb,amsfonts}
\usepackage{algorithmic}
\usepackage{graphicx}
\usepackage{textcomp}
\usepackage{multirow}
\usepackage[nolist]{acronym}
\usepackage{bm}
\usepackage{multicol}
\usepackage{soul,color}
\usepackage{subfig}
\usepackage{booktabs}
\usepackage{lipsum}

\ifCLASSINFOpdf
\else
\fi

\begin{document}

\title{Incorporating Wireless Communication Parameters into the E-Model Algorithm}

\author{Demóstenes Z. Rodríguez,~\IEEEmembership{Senior Member,~IEEE}, Dick Carrillo Melgarejo,~\IEEEmembership{Member,~IEEE}, Miguel A. Ramírez,~\IEEEmembership{Senior Member,~IEEE}, Pedro H. J. Nardelli,~\IEEEmembership{Senior Member}, IEEE,  and Sebastian M\"oller,~\IEEEmembership{Senior Member,~IEEE}

\thanks{This paper is partly supported by Fundação de Amparo à Pesquisa do Estado de São Paulo (FAPESP) under Grant 2015/24496-0 and Grant 2018/26455-8, Conselho Nacional de Desenvolvimento Científico e Tecnológico (CNPq), and Academy of Finland via: (a) ee-IoT project n.319009, (b) FIREMAN consortium CHIST-ERA/n.326270, and (c) EnergyNet Research Fellowship n.321265/n.328869.

D. Z. Rodríguez is with the Universidade Federal de Lavras, Brazil.
(e-mail: demostenes.zegarra@ufla.br).
D. Carrillo and P. H. J. Nardelli are with School of Energy Systems, LUT University, Yliopistonkatu 34, 53850 Lappeenranta, Finland. (e-mail: dick.carrillo.melgarejo@lut.fi, and pedro.nardelli@lut.fi).
M. A. Ramírez is with Department of Electronic Systems Engineering, Escola Politécnica, University of São Paulo, São Paulo 05508-010, Brazil.
(e-mail: maramire@usp.br).
S. M\"oller is with Quality and Usability Lab, Technische Universit\"at Berlin, Germany and Deutsches Forschungszentrum f\"ur K\"unstliche Intelligenz (DFKI), Germany.
(e-mail: sebastian.moeller@tu-berlin.de).
}
}

\markboth{}%
{RODRIGUEZ \MakeLowercase{\textit{et al.}}: Incorporating Wireless Communication Parameters into the E-model Algorithm}
%

\maketitle

\begin{abstract}
Telecommunication service providers have to guarantee acceptable speech quality during a phone call to avoid a negative impact on the users' quality of experience. Currently, there are different speech quality assessment methods. ITU-T Recommendation G.107 describes the E-model algorithm, which is a computational model developed for network planning purposes focused on narrowband (NB) networks. Later, ITU-T Recommendations G.107.1 and G.107.2 were developed for wideband (WB) and fullband (FB) networks. These algorithms use different impairment factors, each one related to different speech communication steps. However, the NB, WB, and FB E-model algorithms do not consider wireless techniques used in these networks, such as Multiple-Input-Multiple-Output (MIMO) systems, which are used to improve the communication system robustness in the presence of different types of wireless channel degradation. In this context, the main objective of this study is to propose a general methodology to incorporate wireless network parameters into the NB and WB E-model algorithms. To accomplish this goal, MIMO and wireless channel parameters are incorporated into the E-model algorithms, specifically into the $I_{e,eff}$ and $I_{e,eff,WB}$ impairment factors. For performance validation, subjective tests were carried out, and the proposed methodology reached a Pearson correlation coefficient (PCC) and a root mean square error (RMSE) of $0.9732$ and $0.2351$, respectively. It is noteworthy that our proposed methodology does not affect the rest of the E-model input parameters, and it intends to be useful for wireless network planning in speech communication services.
\end{abstract}

\begin{IEEEkeywords}
Speech quality assessment, E-model, packet loss, wireless communication, MIMO system.
\end{IEEEkeywords}

\input{acronyms.tex}

\maketitle

\section{Introduction}

\IEEEPARstart{T}HE number of mobile devices has grown rapidly in the last few years, and it will reach $11.6$ billion at the end of $2021$ \cite{Montag2015, CiscoForecast}. Furthermore, in the same year, it is expected that $4G$ Long Term Evolution (LTE) networks will support $53$\% of mobile connections and $79$\%  of mobile data traffic across the world \cite{CiscoForecast}. Based on these forecast data, network operators have to be prepared to manage a high amount of traffic and ensure customer satisfaction, especially in mobile environments. 

In mobile communications systems there are different impairment factors that can affect the users' quality-of-experience (QoE), such as network conditions, environmental acoustic characteristics, and  end-user equipment \cite{raake_terminal, raake_intell}. The speech signal is transmitted through a wireless channel, in which obstacles originate several physical phenomena, such as propagation loss, reflection, diffraction, and scattering.  In addition, the presence of interfering signals and noise degrade the signal quality during its transmission. It is important to note that these impairment characteristics are different from those originated in wired networks. 

Speech quality plays an important role in accomplishing users' expectations of communication services. Cellular network operators perform maintenance tasks, and they need tools to measure the key performance indicators of the network and the speech quality index. At present, there are several speech quality assessment methods. In general, these methods can be classified into subjective and objective ones. There are different kinds of subjective methods, such as conversation opinion tests, interview and survey tests, and listening opinion tests. In the latter tests, assessors score each speech file following a test procedure \cite{itup800}, and an average score called a Mean Opinion Score (MOS) is computed for each speech file. Objective methods use an algorithm to predict an MOS score. They can be classified into three families of models: speech-based, parametric, and hybrid models \cite{sebastian}. Additionally, speech-based models are subdivided into intrusive and nonintrusive methods. Algorithms that use both reference and impaired speech signals are called an intrusive method, and the most representative standardized algorithms are described in the following ITU-T Recommendations: P.862 \cite{itu}, which focuses on narrowband (NB) networks; P.862.2 \cite{itup8622}, which considers wideband (WB) networks; and P.863 \cite{ITUTP863}, which covers networks from NB to fullband (FB) and also presents additional features related to modern communication systems \cite{beerends2013}. On the other hand, algorithms that only use a sole speech signal at the receiving side are known as nonintrusive methods. ITU-T Recommendation P.563 describes a nonintrusive NB-only algorithm, but its scores are not well correlated with subjective test results, especially in lossy channels \cite{polakyy, abareg, pocta_plr, barriac_access, raake_rplr, mittag_plr}. Currently, perceptual approaches for the multidimensional analysis of speech signals considering intrusive (P.AMD) and nonintrusive (P.SAMD) methods are being studied by the ITU-T SG-12 \cite{7776170, pqs2016}. Moreover, in the current literature there are no standardized additional speech quality assessment proposals \cite{PLR2, visqol_1, visqol, 8468124, ietmeu, 8462042, 8463408, 8513822}.

ITU-T Recommendation G.107 \cite{Emodel22} describes a parametric model known as the E-model algorithm, which was initially developed to serve network planning purposes focused on fixed telephone NB networks. Later, the evolution of telecommunication technologies, such as Voice over LTE (VoLTE) networks and the use of proper speech codecs, such as Adaptive Multi-Rate Wideband (AMR-WB) \cite{amrwb}, allowed  transmission of WB speech signals, thereby improving the users' QoE \cite{laura}. Thus, an E-model for WB networks was developed in \cite{Moller}, standardized in G.107.1 \cite{g1071}, and new approaches were introduced in \cite{ieeeaccess, mittag, barriac_swb}. More recently, the E-model for FB networks was standardized in G.107.2 \cite{g1072} enabling to evaluate the speech quality gain of codecs that support FB signals, such as Enhanced Voice Services (EVS) \cite{evs, mittag} and OPUS \cite{opus, pocta_opus} audio codecs.


Despite these updates, the E-model algorithm does not consider parameters related to wireless networks. In addition, the latest wireless network generations incorporate solutions based on the principle of spatial diversity in the radio channel, such as \ac{MIMO} \cite{R5}, in order to improve the data transmission robustness \cite{mimo1,mimo2,mimo3}. In this context, it is important that network planners have specialized tools that take account of wireless degradation characteristics and different communication techniques.



The main contribution of this paper is to propose a general method that incorporates wireless network parameters into the E-model algorithms for NB and WB networks. Thus, the impact of a MIMO system on the speech quality under different wireless channel conditions is studied. This impact is quantified in terms of probability of packet losses; therefore, only the effective equipment impairment factors are modified, specifically $I_{e,\text{eff}}$ and $I_{e,\text{eff,WB}}$ for NB and WB networks, which are defined in \eqref{eq04} and \eqref{eq05}, respectively. The remaining  impairment factors are not affected. In order to evaluate a large variety of transmission quality levels, the additive white Gaussian noise (AWGN) channel model with different background noise intensities is implemented in the wireless communication test scenario. In addition, a MIMO solution with different antenna set configurations is also implemented in the transmission system, being possible to evaluate the effect of this technology on the transmission quality. The speech signal is coded by the AMR or AMR-WB speech codecs, transmitted via a wireless channel, and finally, evaluated at the end point by using the P.863 algorithm. For performance evaluation of the proposed method, subjective tests were carried out. It is worth noting that the proposed methodology was only tested with NB and WB signals. Hence, our proposal intends that the E-model can be useful for network operators and transmission system planners because it considers the techniques used in wireless networks, which are not covered by the E-model algorithm and not included in ITU-T Recommendations G.108 \cite{g108}, G.108.1 \cite{g1081}, and G.108.2 \cite{g1082} either. It is emphasized that the speech quality estimation is only made for transmission planning purposes and not for actual perceptual quality prediction, as stated in \cite{Emodel22, g1071, g1072}.

The remainder of this paper is structured as follows. Section \ref{2} provides an overview of the E-model algorithm.  The speech codecs used in this work are described in Section III. In Section IV, some wireless channel characteristics are presented. The test scenario implementation is described in Section V. In the next section, the proposed methodology is introduced. Section VII presents the experimental results. Finally, conclusions are derived in the last section.


\section{Overview of the E-model Algorithm}
\label{2}

As stated above, the E-model is a computational model used for network planning purposes. It is used to predict the combined impact of various types of degradation on the speech quality during a phone call conversation, such as environmental noise at end-points, attenuation caused by echo, device attenuation at the transmission and reception points, packet losses, and the average delay time in the transmission network. These degradation phenomena are related to different impairment factors, each having an impact on the global speech quality in a linear additive
manner.

The E-model uses a transmission rating scale (R-scale), the ranges of which for NB, WB, and FB are $0$-$100$, $0$-$129$, and $0$-$148$, respectively.  In general, the R-scale is defined by:

\begin{equation}
\label{eq1}
R=R_{0}-I_{s}-I_{d}-I_{e}+A
\end{equation}
where R is the global quality score. $R_0$ represents an ideal scenario without impairments, with default values of $93.2$, $129$, and $149$ for NB, WB and FB networks, respectively. $I_s$ represents all the impairments that occur simultaneously with the voice. $I_{d}$ is related to network delays. $I_e$ is known as the equipment impairment factor, and it represents codec impairments \cite{PLR2}. \textit{A} is the advantage factor; in ITU-T Rec. G.107, \textit{A}  takes, for example, a value of $0$ for a wired network and $20$ for a satellite connection, and in ITU-T Recommendations G.107.1 and G.107.2, \textit{A} is equal to $0$. Moreover, ITU-T Rec. G.107, G.107.1, and G.107.2 define default values in case one of these impairment factors is unknown.

ITU-T Rec. G.107 introduces a function to convert $R$ values into scores on the $5$-point MOS scale. This function is presented in \eqref{eq02}.

\small
\begin{equation}
\label{eq02}
\text{MOS} =
\begin{cases}
1,  R<0 \\
1 + 0.035R + R(R-60)(100-R)(7.10^{-6}), \\
 0<R<100 \\
4.5,  R>100 \\
\end{cases}
\end{equation}
\normalsize

It is pointed out that \eqref{eq02} is only used in the NB context; thus, WB and FB signals have to be converted into an NB signal, for instance by using $R_{NB}=R_{WB}/129$ and $R_{NB}=R_{FB}/148$, for WB and FB signals, respectively.

Because this paper focuses on studying the probability of packet losses, the effective equipment impairment factor relations for NB \cite{Emodel22} and WB \cite{g1071} are presented in \eqref{eq04} and \eqref{eq05}, respectively.

\small
\begin{equation}
\label{eq04}
I_{e,\text{eff}}=I_{e}+(95-I_{e})\frac{\text{Ppl}}{\text{Ppl}/\text{BurstR}+\text{Bpl}}
\end{equation}
\normalsize
where $I_{e}$ is the equipment impairment factor at zero packet-loss, only related to the codec impairment; $Ppl$ is the probability of packet losses, $BurstR$ is the burst ratio, and $Bpl_{WB}$ is the packet-loss robustness factor for a specific codec. 

\small
\begin{equation}
\label{eq05}
I_{e,\text{eff,WB}}=I_{e,\text{WB}}+(95-I_{e,\text{WB}})\frac{\text{Ppl}}{\text{Ppl}+\text{Bpl}_\text{WB}}
\end{equation}
\normalsize
where the variables are the same as those defined in \eqref{eq04} but in the context of WB signals.

In Appendix I of ITU-T Rec. G.113 \cite{itug113}, $I_{e}$ and $Bpl$ values for some codecs or codec families are provided. Similarly, in Appendix IV \cite{itug113anex04}, some $I_{e,WB}$ and $Bpl_{WB}$ values are reported, and in Appendix V \cite{itug113anex05} some $I_{e,FB}$ and $Bpl_{FB}$ values are also defined. 

It is worth noting that ITU-T Rec. P.834 \cite{itup834} and P.834.1 \cite{p8341} describe methodologies based on instrumental models to determine $I_{e}$ and $I_{e,WB}$, respectively. Further, ITU-T Rec. P.833 \cite{itup833} and P.833.1 \cite{itup8331} present methodologies to obtain $I_{e}$ and $I_{e,WB}$, respectively, considering subjective listening quality test results.


\section{Speech codecs used in this work}
\label{11}
This section gives a general overview about the two speech codecs utilized in the present work.

\subsection{AMR speech codec}
The AMR speech codec was initially developed for Global System for Mobile Communications (GSM) networks, and then standardized by the 3GPP \cite{amr}. It has been widely used in GSM and Universal Mobile Telecommunication System (UMTS) networks. This codec uses the Algebraic Code Excitation Linear Prediction (ACELP) algorithm, which is a linear predictive coding (LPC) algorithm based on the code-excited linear prediction (CELP) method.

The AMR is an integrated speech coder with eight operation rates from $4.75$ kbps to $12.2$ kbps, and it is capable of switching its bit rate at each $20$ ms speech segment. In addition, it works with a low bit rate of $1.8$ kbps if Silence Descriptor (SID) frames are continuously transmitted. In general, it consists of different models, a multirate speech coder, a source-controlled rate scheme, a voice activity detector algorithm, a comfort noise generation system, and an error concealment mechanism in case of lost frames in the transmission channels.

\begin{figure}[!ht]
\centering
\includegraphics[scale=0.33]{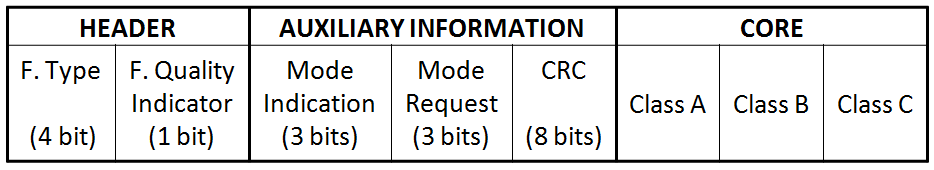}
\caption{Generic AMR frame structure.} 
\label{AMRframe}
\end{figure}

According to \cite{amr22}, each AMR operating mode follows the same frame structure: HEADER, AUXILIARY INFORMATION, and CORE frame as presented in Fig.~\ref{AMRframe}.

As can be observed in Fig.~\ref{AMRframe}, the header includes the frame type field that indicates the use of one of the eight AMR codec modes, and the quality indicator field that informs whether the data in the frame contain errors. The auxiliary information contains the mode indication and request type, and the Cyclic Redundancy Check (CRC) fields that are applied in the most sensitive speech segments. The core considers the speech bits, which are subdivided into three classes according to their subjective importance. Class A bits are the most sensitive, and errors in these bits have a significant impact on the speech quality; thus, any error in these bits typically results in a corrupted speech frame, which should not be decoded without applying an error concealment algorithm. Class A bits are protected by a CRC. Class B bits are more sensitive to errors than Class C bits, but decoding of speech frames with errors is possible without any annoying artifacts. The number of bits used in each class depends on the AMR operating mode. Table~\ref{tab101} presents the main information of each AMR operation mode.

\begin{table}[t]
\caption{Bit composition of the AMR core frame for each operation mode.}
\begin{center}
\scalebox{0.94}{%
\begin{tabular}{c c c c c c c c c c}
\hline
\textbf{\textit{Frame}} & \textbf{\textit{Operating}} & \textbf{\textit{Class}}& \textbf{\textit{Class}}& \textbf{\textit{Class}}& \textbf{\textit{Number}}&  \textbf{\textit{Bit-rate}} \\

\textbf{\textit{Type}} & \textbf{\textit{Mode}} & \textbf{\textit{A}} & \textbf{\textit{B}}& \textbf{\textit{C}}& \textbf{\textit{bits}}& \textbf{\textit{(kbps)}}\\
\hline
\hline
0 & AMR-4.75 & 42 & 53 & 0 & 95 & 4.75 \\ \hline
1 & AMR-5.15 & 49 & 64 & 0 & 103 & 5.15\\ \hline
2 & AMR-5.90 & 55 & 63 & 0 & 118 &5.90\\ \hline
3 & AMR-6.70 & 58 & 76 & 0 & 134 & 6.70\\ \hline
4 & AMR-7.40 & 61 & 87 & 0 & 148 &7.40 \\ \hline
5 & AMR-7.95 & 75 & 84 & 0 & 159 &7.95\\ \hline
6 & AMR-10.2 & 65 & 99 & 40 & 204 &10.2\\ \hline
7 & AMR-12.2 & 81 & 103 & 60 & 244 & 12.2\\ \hline
\end{tabular}%
}
\label{tab101}
\end{center}
\vspace{-4mm}
\end{table}

\subsection{AMR-WB speech codec}
 The AMR-WB codec is specified by the 3GPP in $TS 26.190$ \cite{amrwb}, and it is also described in ITU-T Rec. G.722.2 \cite{g722_2}. It is a speech codec with nine source rates from 6.60 kbps to 23.85 kbps, and a low-rate background noise encoding mode. Similar to the AMR codec, AMR-WB  is capable of switching its bit rate every 20 ms of speech segment, and it is also based on the ACELP algorithm.
 
 The AMR-WB frame structure of each operating mode is the same as for the AMR codec presented in Fig.~\ref{AMRframe}, and also the number of bits of the header and auxiliary information frames are the same. Because of the different bit rates of each operation mode, their core frame lengths are different. Furthermore, Class C bits are not considered in the AMR-WB codec \cite{amr22} as can be observed in Table~\ref{tab101AMRWB}.
 
 \begin{table}[!htbp]
\caption{Bit composition of the AMR-WB core frame for each operation mode.}
\begin{center}
\scalebox{0.94}{%
\begin{tabular}{c c c c c c c c c c}
\hline
\textbf{\textit{Frame}}  & \textbf{\textit{Class}}& \textbf{\textit{Class}}&  \textbf{\textit{Number}}&  \textbf{\textit{Bit-rate}} \\

\textbf{\textit{Type}}  & \textbf{\textit{A}} & \textbf{\textit{B}}& \textbf{\textit{bits}}& 
\textbf{\textit{(kbps)}}\\
\hline
\hline
0  & 54 & 78  & 132 & 6.60 \\ \hline
1  & 64 & 113  & 177 & 8.85\\ \hline
2  & 72 & 181 & 181 & 12.65\\ \hline
3  & 72 & 76  & 213 & 14.25\\ \hline
4  & 72 & 87  & 245 & 15.85 \\ \hline
5  & 72 & 84  & 293 & 18.25 \\ \hline
6  & 72 & 99  & 325 & 19.85 \\ \hline
7  & 72 & 103 & 389 & 23.05\\ \hline
8  & 72 & 103 & 405 & 23.85\\ \hline
\end{tabular}%
}
\label{tab101AMRWB}
\end{center}
\end{table}

 The AMR-WB codec uses a sample rate of $16$ kHz, offering an improved speech quality in relation to the AMR-NB codec. It is used  in current speech communication networks because it achieves a reasonable performance in different network conditions, providing high-quality speech communications. 
The AMR-WB performance assessment with music signals is only satisfactory at the highest bit rate, according to listening quality tests \cite{amrperform}.

\section{Wireless Channel Characteristics and MIMO Systems}
\label{3}

To improve the quality of transmission systems, network architecture researchers need to understand the key performance parameters of wireless channels that influence the choice of transmission techniques. Speech quality is a relevant parameter in voice communication systems, and it depends on the transmission channel conditions.

In a wireless communication scenario, usually the end-points of a communication are in different places and have many obstacles between them, such as city environments. Thus, it is probable that there is non-line-of-sight (NLOS) between the sender and the receiver; the communication is established by the scattering of the waves or by diffraction around the objects \cite{siswire}, which in most of the cases are destructive components on the main path component. For instance, an impairment characteristic in wireless channels is fading \cite{6517180,5301913,6804500}, which is due to several conditions, such as different signal paths, channel frequency variation, signal power attenuation, and relative movement between sender and receiver. As a consequence, the amplitude and frequency of the transmitted signal varies randomly \cite{gibson2002communications}. There are different stochastic models that are frequently used in the literature, some of them consider that one of the paths is much stronger than others \cite{rician1}, and other models consider that there is no dominant propagation along a line-of-sight (LOS), and the signal spreads around various obstacles until it reaches the receiver \cite{rayle1}. Additive white Gaussian noise (AWGN) channel model is commonly used to represent different background noise of the channel under study, and the noise added by transceiver components at the receiver \cite{745083422}. The AWGN represents a simpler mathematical model that is useful for gaining insight into the underlying behavior of a communication system, and in general, it is used to test new transmission solutions or network architectures due to its simplicity \cite{awgn222}.

%


In wireless communications, many solutions are used to improve the system performance. MIMO systems provide a cost-effective approach to high-throughput wireless communications. This concept has been studied for a long time to support wireless systems.
It was used in many cellular technologies, such as the \ac{UMTS} and \ac{HSDPA} based \acp{3G}; LTE and LTE-Advanced (LTE-A)  based \ac{4G}; and today, it is a key feature in initial deployments of \ac{5G}  when a large number of antennas are considered. 
%

\begin{figure}[!htbp]
\centering
\includegraphics[scale=0.48]{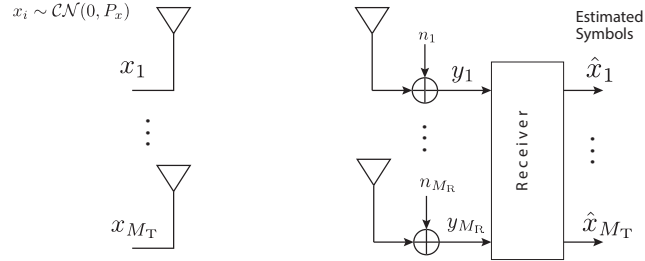}
\caption{General \ac{MIMO} system model with $M_\text{T}$ antennas transmitting and $M_\text{R}$ antennas receiving.} 
\label{mimo_system}
\end{figure}

The significant contributions of \ac{MIMO} systems are related to the system capacity and transmission robustness.
%
When the objective is to use the \ac{MIMO} technology in the wireless communication channel to introduce diversity gain for a better transmission performance, a general \ac{MIMO} architecture could be considered, where the processing of each element of the model is described on a symbol-by-symbol basis.
Thus, a general \ac{MIMO} system depicted in Fig. \ref{mimo_system} is mostly characterized by $M_\text{T}$ and $M_\text{R}$ antennas on the transmitted and receiver sides, respectively.

Considering that each transmitted package is encoded independently and transmitted in different antennas simultaneously, the \ac{MIMO} system is based on $M_\text{T}$ antennas on the transmitted side, and $M_\text{R}$ antennas on the receiver side.
The system could be modeled by: 
\begin{equation}
\label{eq:mimo_general_model}
    \bm{y} = \bm{H} \bm{x}+\bm{n}
\end{equation}
where the packets transmitted by $M_\text{T}$ are denoted by a vector column $\bm{x} \in \mathcal{C}^{ M_\text{T} \times 1 }$ with elements $x_i$, where $i=0,\cdots,M_\text{T}$, and the received packets are also denoted by a vector $\bm{y} \in \mathcal{C}^{ M_\text{R} \times 1 }$ with elements $y_j$, where $j=0, \cdots, M_\text{R}$. It is also important to mention that each stream of data is allocated to the same power $P_x=\Bar{P}/M_\text{T}$ and $\Bar{P}$ is the total transmit power over all the transmit antennas.
The packets are subjected to the adverse effects of the propagation channel $\bm{H} \in \mathcal{C}^{M_\text{R}\times M_\text{T}}$, and the additive noise at the receiver is modeled by $\bm{n} \in \mathcal{C}^{M_\text{R} \times 1}. $ 

%
%
The design of a \ac{MIMO} strategy is directly related to the knowledge about this channel, this information being defined as \ac{CSI}.
%
However, when the \ac{CSI} is not available, the second-order statistics and long-term properties of the channel are considered.
This availability is also an important aspect when choosing the \ac{MIMO} scheme.

To encode the signals that will be transmitted via MIMO systems, there are different proposed encode algorithms. Some of these algorithms are the Block Layered Space-Time (BLAST) \cite{6770094}, Space-Time Trellis Codes (STTCs) \cite{825818, 4339800}, Space-Time Block Codes (STBCs) \cite{730453}, Unitary Space-Time Codes \cite{1374943} among others. Orthogonal STBCs (OSTBC) \cite{proakiss} admits a simplified decoding at the receiver side. The MIMO performance, and as a consequence the transmission quality, depends on the implemented encode algorithm. The OSTBC system only needs the \ac{CSI} at the reception side, and it also has a low spatial correlation allowing the system to be more robust against fading degradation.

\section{Test Scenario Implementation}
\label{4}

\begin{figure}[!ht]
\centering
\includegraphics[scale=0.39]{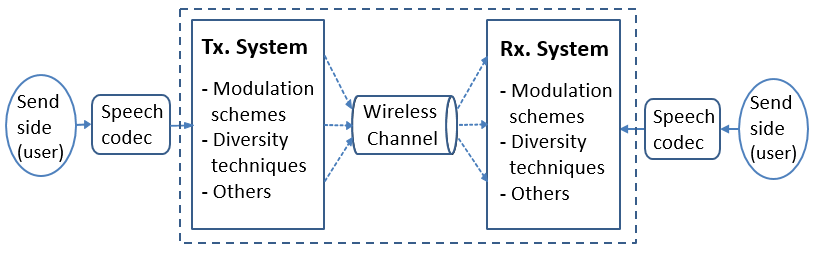}
\caption{High-level representation of some techniques used in wireless communication systems.}
\label{scenariow}
\end{figure}

In the case of a wireless network, the transmission and reception systems use different techniques from those used in wireline networks, and determination of their impact on the transmission quality can be useful for network operators and transmission planners. Fig.~\ref{scenariow} presents a basic high-level representation of some techniques used in the transmitter and receiver of a wireless communication system.

Modulation schemes play an important role in wireless communication systems; for example, depending on the modulation scheme used in the transmission system, different data rates can be achieved, and each scheme has a different response at a specific channel background noise level. This can be observed in the packet loss rate obtained by each modulation scheme under the same transmission channel conditions.

\begin{figure*}[!ht]
\centering
\includegraphics[scale=0.52]{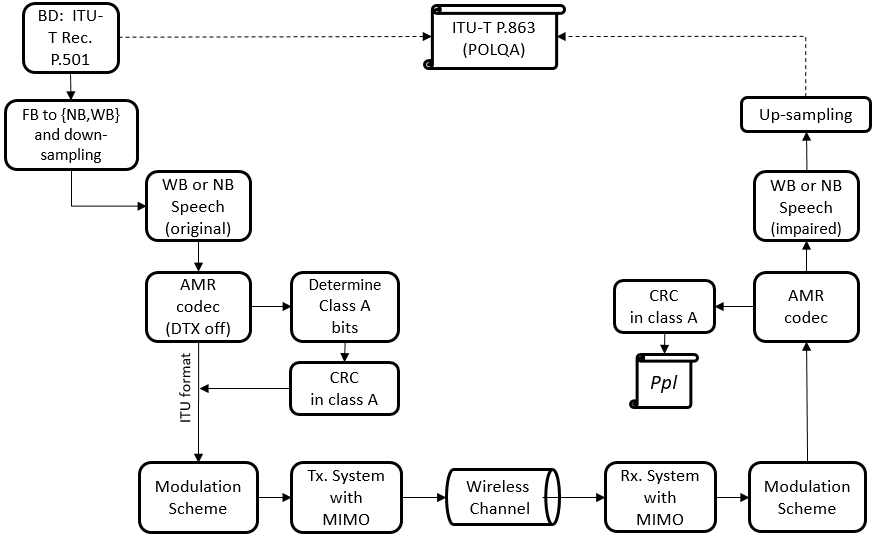}
\caption{Block diagram of the implemented test scenario.} 
\label{scenario}
\end{figure*}

There are various approaches to implement diversity. There are three main types of diversity: time diversity, frequency diversity and space diversity. Diversity over time can be obtained by coding and interleaving; the information is coded and dispersed over time in different coherence periods with the objective that each segment of the information suffers independent fades. Similarly, the frequency diversity can be applied if the channel is frequency-selective.
A solution to enhance the communication performance is to establish multiple signal paths, each of which fades independently, making sure that reliable communication is possible achieved as long as one of the paths is strong enough. In general, the diversity techniques transmit the information in different transmission channel conditions, with the aim of decreasing the data error rate at the receiver \cite{Tse}. Space diversity is implemented in a channel with multiple transmit or receive antennas. Because diversity techniques provide important advantages, a wireless communication system typically uses several types of diversity \cite{Tse}. A \ac{MIMO} system defined in  (\ref{eq:mimo_general_model}) is an example of a space diversity technique that is commonly adopted and used in wireless transmission systems. 

In order to study the impact of wireless channel degradation on speech quality, a wireless network simulator was built considering each element described above. The block diagram of the implemented test scenario is presented in Fig.~\ref{scenario}.

In the test scenario, test signals extracted from Annex C of ITU-T Recommendation P.501 \cite{itup501} were used, which are originally FB speech sequences with a sampling frequency of $48$ kHz. In all samples, the amount of active speech is greater than $3.0$ s as recommended in \cite{itup8631}. 
As our focus is to evaluate the NB and WB scenarios, two band-pass filters \cite{itup341} were applied to the original signals. Further, the speech level was equalized to $-26$ dBov using \cite{itup191}, and a down-sampling process was applied to obtain a sampling rate of $16$ kHz. 
In our simulation tests, four unimpaired speech files were considered (two male and two female speakers).

In the tests, the AMR-WB codec was considered because it is the most commonly used one in current $4G$ networks, and the AMR codec is used in existing $2G$ and $3G$ networks. 

As presented in Fig.~\ref{AMRframe}, the AMR-WB header includes the FQI field, which indicates if the data in the frame contain errors, and its status can be used by error concealment algorithms implemented in the decoder. In the test scenarios, the FQI value was determined by using the information given by the CRC code implemented in the AMR and AMR-WB codecs.

It is noteworthy that the CRC implementation is not available in \cite{amr} and \cite{amrwb}. In order to quantify the actual $Ppl$ value in each simulation, a CRC code was implemented following the instructions given in \cite{amr2} and \cite{amr22} for AMR and AMR-WB, respectively. It is pointed out that the CRC code is computed only considering Class A bits. Moreover, the output bits of the AMR and AMR-WB codecs are ordered according to their subjective importance, which is useful for error protection purposes. The AMR and AMR-WB core frame bit distributions are introduced in Annex B of \cite{amr2} and \cite{amr22}, respectively.

The modulation schemes used in the test scenarios are BPSK, QPSK, and QAM (QAM-16, QAM-32, QAM-64 and QAM-256). The PSK and QAM modulation schemes are used in the subcarriers of the OFDM. 

In the implemented MIMO system, the signal can be transmitted and received by one to four antennas. In the test scenarios, the MIMO antenna configurations 
$(M_\text{T}, M_\text{R})$, $2\times 2$, $3\times 3$, and $4\times 4$ as well as the Single-Input-Single-Output (SISO) were applied. The MIMO system uses the Orthogonal Space-Time Block Code (OSTBC) encoder. The OSTBC maps the symbols coming from previous phases in the transmitter in order to generate temporal vectors containing the symbols that will be sent by each of the antennas. At the receiver, the OSTBC combines the received signals by all the antennas to extract the information from the symbols encoded in the transmitter \cite{ostbc}. 

The transmission channel block corresponds to the AWGN channel model, in which different noise intensities are applied. In order to restrict the number of simulation scenarios, the AWGN was configured using SNR values from $0$ dB to $30$ dB, with steps of $1$ dB. Thus, in this work, the channel model is only represented by an SNR value. At the reception side, the signals are demodulated and decoded.

Finally, the resulting impaired speech files were up-sampled to $48$ kHz and then evaluated by the P.863 algorithm, thereby obtaining a MOS value for each test scenario. It is important to note that the filters applied in the original FB signals \cite{itup341, itup8631} avoided any aliasing effect; thus, the sampling rate conversion did not present an audible degradation impact on speech quality.
 
 \section{Proposed methodology that permits the use of wireless network parameters into the E-model algorithm}
\begin{figure}[!htpb]
\centering
\includegraphics[scale=0.38]{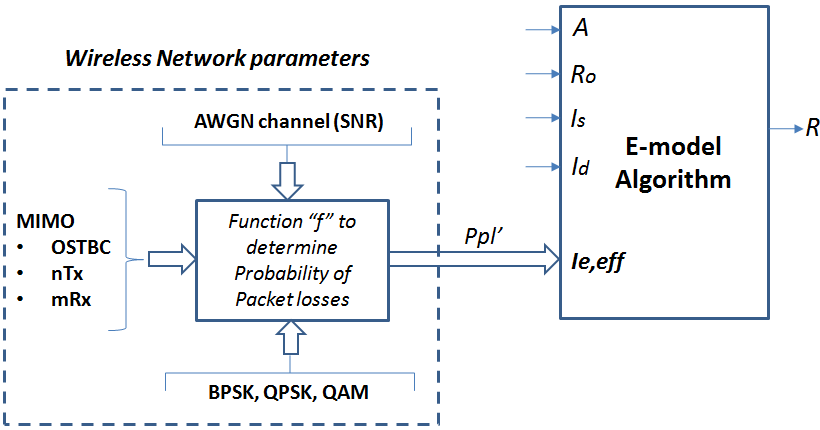}
\caption{High-level representation of the proposed method.} 
\label{metodo}
\end{figure}

The high-level representation of the proposed methodology to incorporate wireless networks parameters into the E-model algorithm is presented in Fig.~\ref{metodo}. 

As can be observed in Fig.~\ref{metodo}, the proposed methodology aims at determining an estimated $Ppl$ ($Ppl'$) parameter using only wireless network information. Thus, it only has an impact on the $I_{e,eff}$ impairment factor, while other E-model variables are not affected. 

\begin{figure*}[!ht]
\centering
\includegraphics[scale=0.38]{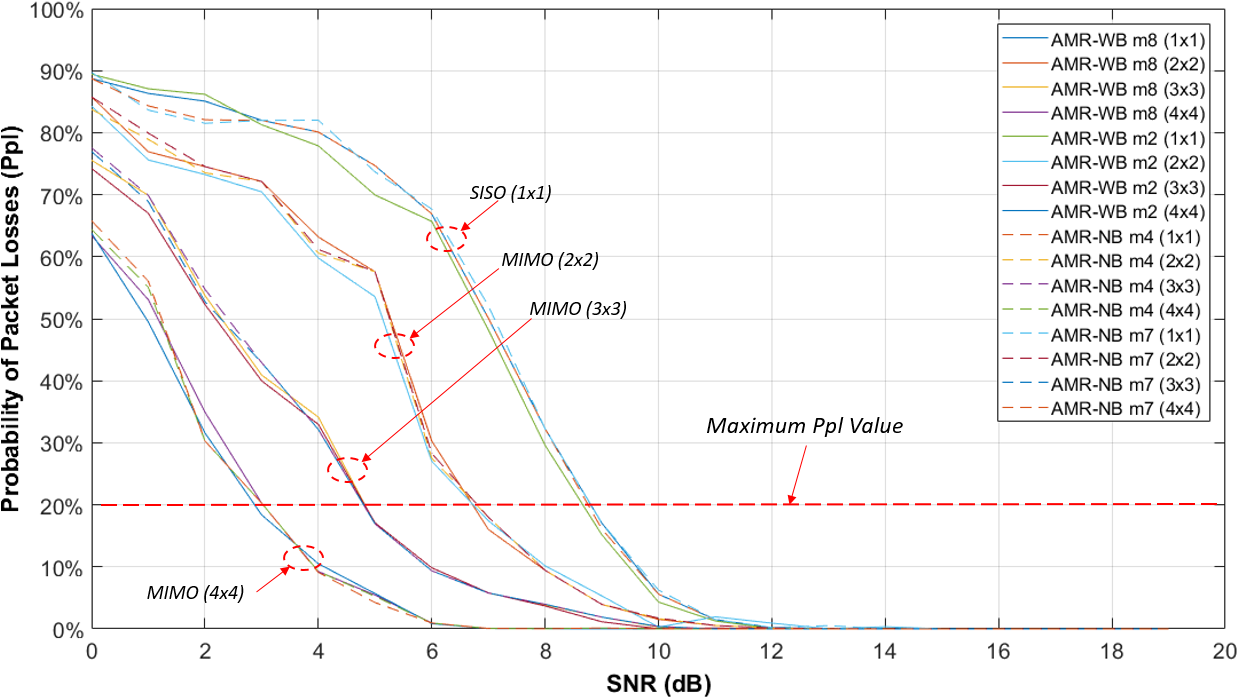}
\caption{Evaluation of the AMR (AMR-NB) and AMR-WB speech codecs under different wireless channel conditions considering the Ppl parameter and using the QPSK modulation.} 
\label{ppl_amr}
\end{figure*}

Firstly, in order to determine the impact of the speech codec on the $Ppl$ values, two different operation modes of the AMR and AMR-WB speech codecs were used in the test scenario under the same wireless channel conditions, specifically an AWGN channel model with different SNR intensities from $0$ dB to $30$ dB. In addition to the SISO scenario, three MIMO antenna configurations were used. To obtain more representative results, each network scenario was run for $60$ times.  Thus, it is possible to observe the $Ppl$ values for each speech codec. Fig.~\ref{ppl_amr} shows the experimental results for the AMR modes $4$ and $7$ and the AMR-WB modes $2$ and $8$, in combination with the QPSK modulation scheme.

In this initial test, a large range of $Ppl$ values were considered, but according to ITU-T Rec. G.107, the permitted range of random $Ppl$ is from $0$ to $20$\% (dashed horizontal line in Fig. 6), which is closer to actual network conditions.  Fig.~\ref{ppl_amr} shows that the distribution of $Ppl$ values is almost the same for every speech codec. Thus, it can be concluded that $Ppl$ depends mainly on the transmission channel conditions and the MIMO antenna system configuration.

In the case of wireless transmission, the $Ppl$ can be estimated by a function $f$, which depends on one or more parameters of the implemented diversity technique, modulation scheme, and wireless channel model. Then, in a general wireless context, $Ppl'$ is defined as

\begin{equation}
\label{eq7}
Ppl^\prime= f(D_t, M_s, W_c)
\end{equation}

where $Ppl'$ is the estimated $Ppl$ value, $D_t$ represents the parameters of the diversity technique, $M_s$ corresponds to the modulation scheme type, and $W_c$ represents the wireless channel parameter set.

Then, for a wireless network, the estimated $I_{e,eff}'$ for NB networks can be determined by

\begin{equation}
\label{eq77}
I_{e,eff}'= I_{e}+(95-I_{e})\frac{f(D_t, M_s, W_c)}{f(D_t, M_s, W_c)/BurstR + Bpl}
\end{equation}

Furthermore, for WB networks, the $I_{e,eff,WB}$ can be calculated by

\begin{equation}
\label{eq78}
I_{e,eff,WB}'= I_{e,WB}+(95-I_{e,WB})\frac{f(D_t, M_s, W_c)}{f(D_t, M_s, W_c) + Bpl_{WB}}
\end{equation}

Later, by using \eqref{eq77} and \eqref{eq78} and the remaining impairment factors of the corresponding E-model algorithm, the R-score can be determined. As stated before, the $I_{e}$ and $Bpl$ values for NB codecs are taken from Appendix I of ITU-T Rec. G.113 \cite{itug113}, and Appendix IV \cite{itug113anex04} of the same recommendation provides the $I_{e,WB}$ and $Bpl_{WB}$ values. Thus, it is possible to estimate quantitatively, using the R-scale, the effect of any wireless network parameter on the speech quality.

For example, if the $D_t$ is a MIMO system, in which the evaluated parameters are only the number of transmission $(nTx)$ and reception $(mRx)$ antennas, the $W_c$ model is AWGN, whose background noise is only represented by SNR (dB), and finally, the $M_s$ is QPSK; thus, the $\text{Ppl}^\prime$ can be determined by $\text{Ppl}^\prime = f((\text{nTx},\text{mRx}), \text{QPSK}, \text{SNR})$.    

Next, numerical values of the $Ppl$ parameter for each wireless network condition need to be obtained. To accomplish this objective, experimental tests are required. Thus, a network scenario has to be implemented, in which each of the network parameters can be controlled, and the actual $Ppl$ values can be determined. In this study, the network scenario introduced in Fig.~\ref{scenario} is used.
Further, the implemented test scenario has to be validated to guarantee the accuracy of the obtained $Ppl$ values. For example, the resulting impaired speech samples of each scenario can be evaluated by instrumental methods, such as the algorithm described in ITU-T Recommendation P.863, or subjective tests to obtain speech quality scores. On the other hand, the NB or WB E-model algorithm can be used considering the network parameters, including $Ppl$ values, to obtain the R-scores. Then, a high correlation between both of those results should be obtained. 

In order not to restrict $\text{Ppl}^\prime$ to only the SNR values used in the experimental test scenarios, the function to fit the $\text{Ppl}^\prime$ results, introduced in \eqref{eq7}, is determined and used for planning purposes. For instance, in the case of a MIMO system, it will be possible to estimate the $\text{Ppl}^\prime$ value for each MIMO antenna configuration; thus, the speech quality can be estimated by using the corresponding E-model algorithm.

\section{Experimental Evaluation and Results}
\label{6}
In this section, the validation of test scenario results, the proposed function to determine \textit{Ppl'} values, and the performance of the proposed methodology using subjective tests are treated. At the end of this section final discussions are described.

\subsection{Test scenario validation}

In order to test different network conditions, several test scenarios were implemented, considering different $M_s$ and MIMO antenna configurations.  Table~\ref{tab0006} presents the configuration parameters used in each test scenario.

\small
\begin{table}[!htbp]
\caption{Parameters used in the test scenarios.}
\begin{center}
\scalebox{0.98}{%
\resizebox{\columnwidth}{!}{%
\begin{tabular}{c c }
\hline
\textbf{\textit{Network Parameter}} &  \textbf{\textit{Description}} \\
\hline
\hline
$M_s$ & BPSK, QPSK, QAM (16, 32, 64, 256)   \\ 
$D_t$ (Antenna configuration) & $(1\times1) , (2\times2), (3\times3), and  (4\times4)$   \\ 
$W_c$ (SNR) & 0, 1. 2 ... 28, 29, 30 (dB)   \\ 
Speech Codec & AMR (modes: 4, 7) and AMR-WB (modes: 2, 8)    \\ 
\hline
\end{tabular}}%
}
\label{tab0006}
\end{center}
\vspace{-3mm}
\end{table}
\normalsize
From Table~\ref{tab0006}, a total of $2,976$ different test network scenarios can be computed  ($6$ $M_{s}$, $31$ channel degradation, $4$ speech codecs, and $4$ antenna configurations). An SNR range from $0$ dB to $30$ dB was chosen to guarantee that all network configurations used in test scenarios permit to reach the highest and lowest speech quality scores, for example, to reach a maximum quality score, using $M_{s}$ = QAM-256, SNR values close to $30$ dB are necessary.  In addition, to have more representative results of each scenario, $4$ unimpaired speech samples (two male and two female speakers) were used as input, and  each scenario was simulated for $60$ times. Thus, a total of $714,240$ impaired speech samples were created from all the test scenarios. 

\begin{figure}[!ht]
\centering
\includegraphics[scale=0.34]{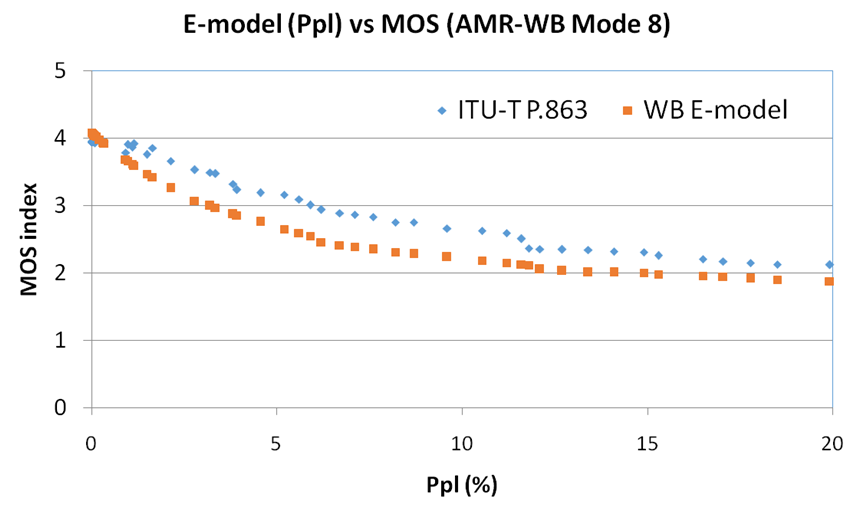}
\caption{Performance validation of the implemented test scenario using AMR-WB with mode operation $8$ and a SISO configuration.} 
\label{perform}
\vspace{-3mm}
\end{figure}

As the first step, the test scenario performance validation was evaluated only with respect to an SISO implementation [$D_t$:$(1\times1)$], and considering all possible variations of the remaining parameters described in Table~\ref{tab0006}. This validation was performed using two objective measures, the ITU-T Rec. P.863 results and E-model scores. To this end, the actual $Ppl$ value was determined in each test scenario by using the CRC code implementation \cite{p8341}. The R-score quality was determined by using \eqref{eq04} for the NB scenarios or \eqref{eq05} for the WB scenarios, and then \eqref{eq1}. Later, these R-scores were converted to the 5-point MOS scale using (2), and then they are compared with the ITU-T Rec. P.863 results. Fig.~\ref{perform} shows the performance results of the AMR-WB speech codec mode $8$ and considering a SISO implementation.

Based on the results obtained by the E-model and P.863 algorithms presented in Fig.~\ref{perform}, the Pearson correlation coefficients (PCC) and the root mean squared error (RMSE) are determined, reaching $0.982$ and $0.168$, respectively. It can also be observed that the values computed using the E-model algorithm are slightly lower than those reached by ITU-T Rec. P.863, which is recommendable for planning purposes because network planners can provide more resources to guarantee an acceptable transmission quality. Similarly, the other speech codecs were evaluated, and the results are presented in Table~\ref{tab0007}.

\begin{table}[!ht]
\caption{Validation results of the network scenario implementation.}
\begin{center}
\scalebox{0.62}{
\resizebox{\columnwidth}{!}{%
\begin{tabular}{c c c}
\hline
\textbf{\textit{Speech codec - Mode}} &  \textbf{\textit{PCC}} & \textbf{\textit{RMSE}}\\
\hline
\hline
AMR Mode 4 & 0.981 & 0.195 \\ 
AMR Mode 7  & 0.975 & 0.156  \\ 
AMR-WB Mode 2 & 0.979 & 0.173  \\ 
AMR-WB Mode 8 &  0.982  & 0.168  \\ 
\hline
\end{tabular}
}
}
\label{tab0007}
\end{center}
\end{table}

It is emphasized that the results presented in Table~\ref{tab0007} show that the CRC code implementation to determine the actual $Ppl$ values is highly reliable. 

\subsection{Performance validation of the proposed function to determine Ppl' values}

\begin{figure}[!ht]
\centering
\includegraphics[scale=0.34]{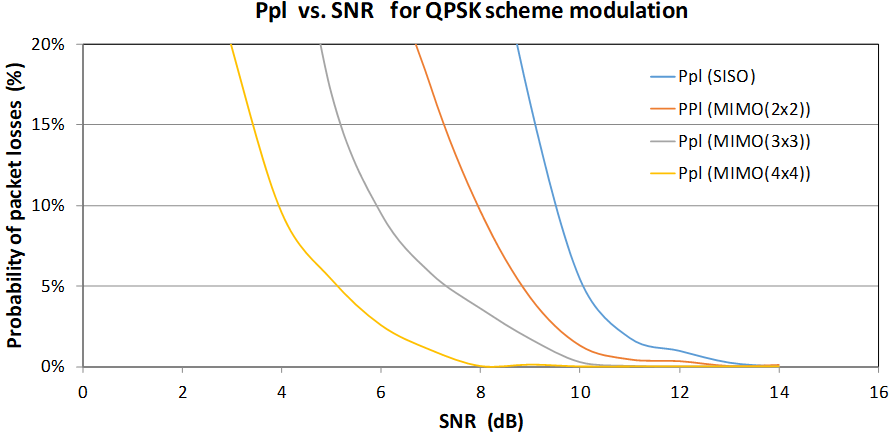}
\caption{Ppl values (\%) for $M_s$ = QPSK using a SISO and three MIMO antenna configurations.} 
\label{QPSK_ALL}
\vspace{-4mm}
\end{figure}

For simplicity and better explanation of the results, only a case study that corresponds to a specific network configuration was used to present the results obtained by the proposed methodology. This configuration considers the SISO and the MIMO antenna set (n$T_x$, m$R_x$) equal to (2,2), (3,3), and (4,4); the encode MIMO algorithm is OSTBC; the modulation scheme is QPSK; the transmission channel is AWGN; and the $Ppl$ values is limited to 20\%. Fig.~\ref{QPSK_ALL} presents the results for these specific test scenarios. Note that these results correspond to all speech codecs and their mode operations presented in Table~\ref{tab0006}.


Based on the results presented in Fig.~\ref{QPSK_ALL}, the function {\it f} introduced in \eqref{eq7} was determined. This function was used to calculate the estimated {\it Ppl (Ppl’)}. Different mathematical relations were tested, and the function that best fits the results for each MIMO antenna configuration was the power function, as presented in 

\begin{equation}
\label{eq10}
Ppl'= aSNR^b+c
\end{equation}
where a, b, and c are the coefficients of the power function, and SNR is expressed in dB. 

Table~\ref{tab555} presents the coefficient values of $a, b,$ and $c$ introduced in \eqref{eq10} for each MIMO antenna configuration.

\small
\begin{table}[!htbp]
\caption{Coefficient values of the function used to determine \textit{Ppl'} using QPSK.}
\begin{center}
\scalebox{0.85}{%
\resizebox{\columnwidth}{!}{%
\begin{tabular}{c c c c c}
\hline
\textbf{\textit{$M_s$}} &  \textbf{\textit{(n$T_x$, m$R_x$)}} & \textbf{\textit{a}} & \textbf{\textit{b}} & \textbf{\textit{c}}\\
\hline
\hline
\text{QPSK} & (1,1) & $8395x10^{6}$ & -11.2 & -0.0004646 \\ 
\text{QPSK}  & (2,2) & 43900 & -6.434 & -0.0001633 \\ 
\text{QPSK} & (3,3) & 60.61 & -3.651 & -0.0009027  \\ 
\text{QPSK} &  (4,4)  & 12.12 & -3.722 & -0.0002535 \\ 
\hline
\end{tabular}}%
}
\label{tab555}
\end{center}
\end{table}
\normalsize

\begin{figure}[!ht]
\centering
\includegraphics[scale=0.34]{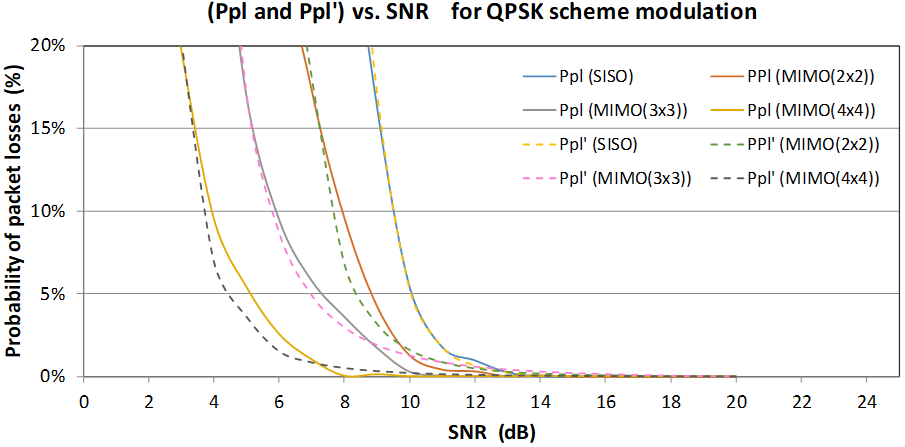}
\caption{ {\it Ppl'} values obtained using the proposed function ($Ppl'= aSNR^b+c$), and actual {\it Ppl} values, considering MIMO system transmission and QPSK modulation.} 
\label{shapef}
\end{figure}

Fig.~\ref{shapef} depicts the {\it Ppl’} results obtained by using \eqref{eq10} and applying the values of the constants $a, b,$ and $c$ presented in Table~\ref{tab555}. In this same figure the {\it Ppl} values are also presented for comparison purposes.

The coefficient of determination ($R^2$) and the RMSE are used as the performance metrics to determine the performance validation results of \eqref{eq10} for each antenna set configuration {\it (nTx, mRx)}. Thus, the $R^2$ and the RMSE are calculated considering the \textit{Ppl} and \textit{Ppl'} values, and they are presented in Table~\ref{tab666}.  


\small
\begin{table}[!htbp]
\caption{Performance validation results of the function used to determine \textit{Ppl’} in relation to actual \textit{Ppl} for each antenna set configuration.}
\begin{center}
\scalebox{0.56}{%
\resizebox{\columnwidth}{!}{%
\begin{tabular}{c c c }
\hline
\textbf{\textit{(n$T_x$, m$R_x$)}} &  \textbf{\textit{$R^2$}} & \textbf{\textit{RMSE}} \\
\hline
\hline
SISO(1,1) & 0.9985& 0.001554  \\ 
MIMO(2,2)  & 0.9963 & 0.002394  \\ 
MIMO(3,3) & 0.9973 & 0.002069   \\ 
MIMO(4,4) &  0.9964 & 0.006245 \\ 
\hline
\end{tabular}}%
}
\label{tab666}
\end{center}
\vspace{-4mm}
\end{table}
\normalsize

It can be observed from Table~\ref{tab666} that the proposed function presented in \eqref{eq10} to calculate \textit{Ppl’} values reached \textit{$R^2$} scores higher than $0.996$, and RMSE scores lower than $0.0063$ in relation to actual \textit{Ppl} values, in all the antenna set configurations. With these reliable results, the next step is to use \textit{Ppl’} into the appropriated E-model algorithm, called E-model (\textit{Ppl’}).

\subsection{Performance comparison results  between the E-model (\textit{Ppl’}) and the E-model (\textit{Ppl}) }

The relation \eqref{eq10} is used to estimate the $I_{e,eff}$ and $I_{e,eff,WB}$. To accomplish this task, the parameters $I_{e}$ and $Bpl$  for NB and WB context of each speech codec mode operation need to be known. Once the $I_{e,eff}$ and $I_{e,eff,WB}$ values are estimated, the R-scores are predicted using the appropriated E-model  (\textit{Ppl’}).  For instance, if the AMR-WB speech codec mode $2$ is evaluated, the $I_{e,WB}$ and $Bpl_{WB}$ values from \cite{itug113anex04}, and the coefficient values from Table V are considered. Then, the estimated $I_{e,eff,WB}$ -- denoted as $I_{e,eff,WB}$’ -- is obtained by the following relation:

\small
\begin{equation}
\label{eq11}
I_{e,eff,WB}'= I_{e}+(95-I_{e,WB})\frac{aSNR^b+c}{(aSNR^b+c) + Bpl_{WB}}
\end{equation}
\normalsize

Later, the R-score can be calculated using \eqref{eq1} in WB context. For performance evaluation, the results obtained by the WB E-model using both the \textit{Ppl} and the \textit{Ppl’} values using \eqref{eq10} are compared. The performance assessment of the AMR-WB mode $2$ for the four antenna configurations used in the tests reached a PCC and an RMSE of $0.9846$ and $2.6841$, respectively. These results demonstrate that the estimated R-scores based on \textit{Ppl’} and the actual R-scores are highly correlated; therefore, the proposed methodology works properly if the speech codec information is available.

The same procedure to compute R-scores was applied to the rest of the speech codec modes using the appropriated E-model algorithm. In the case of AMR mode 4, the $Bpl$ value reported in \cite{5759942} is considered, because it is not available in \cite{itug113}. The performance results regarding the comparison between the E-model (\textit{Ppl’}), and the E-model (\textit{Ppl}) are presented in Table~\ref{tab777}. Note that these results correspond to all the antenna set configurations and the QPSK modulation scheme. 

\small
\begin{table}[!htbp]
\caption{Performance validation results of the E-model (\textit{Ppl’}) in relation to the E-model (\textit{Ppl})  considering all the antenna set configurations and each speech codec mode.}
\begin{center}
\scalebox{0.70}{%
\resizebox{\columnwidth}{!}{%
\begin{tabular}{c c c c}
\hline
\textbf{\textit{$M_s$}} & \textbf{\textit{Speech codec}} &  \textbf{\textit{PCC}} & \textbf{\textit{RMSE}} \\
\hline
\hline
QPSK & AMR Mode 4 & 0.9795& 2.4453  \\ 
QPSK &AMR Mode 7  & 0.9871 & 2.7822  \\ 
QPSK &AMR-WB Mode 2 & 0.9846 & 2.6841  \\ 
QPSK &AMR-WB Mode 8 &  0.9764 & 2.1594  \\ 
\hline
\end{tabular}}%
}
\label{tab777}
\end{center}
\vspace{-4mm}
\end{table}
\normalsize

As can be observed from Table~\ref{tab777}, the E-model (\textit{Ppl’}) results are reliable.  Hence, if the MIMO configuration, the wireless channel conditions, the speech codec parameters, and the modulation scheme are known, the R-score can be estimated following our proposed methodology. To this end, only the $I_{e,eff}$ or $I_{e,eff,WB}$ is considered, the remaining of E-model parameters are not modified by our proposal.

The same methodology presented in this case study using QPSK modulation was applied to the other modulation schemes. The coefficient values introduced in \eqref{eq10} that correspond to each $M_s$ and their performance validation results are presented in the appendix of this document.

Additionally, performance validation tests were performed using different wireless channel parameters. The parameters were: $W_c$ with SNRs equal to $1.5$ dB, $4.5$ dB, $8.5$ dB, $12.5$ dB, $16.5$ dB, $20.5$ dB, and $24.5$ dB; $M_{s}$ corresponding to QPSK, QAM-16, and QAM-256; AMR modes $3$ and $7$, AMR-WB $2$ and $8$; and antenna configurations MIMO(2x2) and MIMO(4x4). Each test scenario was repeated for $60$ times and two unimpaired speech samples were used in each one. The speech quality was estimated in each of the $20,160$ scenarios by the appropriated E-model using ${Ppl'}$ and ${Ppl}$ values. Based on the experimental results, the comparison between E-model (${Ppl'}$) and E-model (${Ppl}$) reached a PCC and an RMSE of $0.979$ and $3.227$, respectively.

These reliable results, presented in this section so far, were obtained through objective methods. A performance evaluation using subjective tests is presented in the next subsection.

\subsection{Global performance assessment of the proposed methodology trough subjective tests}

Finally, subjective tests of speech quality assessment were carried out in a controlled laboratory environment. These tests were needed because the parameters related to wireless networks are not addressed in the current E-model algorithms, and the proposed methodology had to be exhaustively tested. 
To this end, a total of 640 impaired speech samples were used in the listening audio quality tests. These speech samples were chosen from the ones generated by using the network parameters presented in Table~\ref{tab0006}; thus, for each sample, the network scenario configuration and the MOS values given by the ITU-T Tec. P.863 are known.

In order to have a test material containing speech samples with different perceptual quality levels, a homogeneous distribution according to the quality score was done, using the 5-point MOS scale. Thus, the MOS given by ITU-T Rec. P.863 were used to distribute the samples. Furthermore, all the parameters presented in Table~\ref{tab0006} were used and evenly distributed.
In the subjective tests, a total of 104 subjects participated, including 45 females and 59 males, aged between 18 and 56 years. They were volunteers, did not receive any financial compensation, and none of them reported any hearing problem or experience in the speech quality assessment task. 
The tests were conducted according to the Absolute Category Rating (ACR) method described in ITU-T Rec. P.800 \cite{itup800} during a period of 11 weeks, and in this period, the test room and audio equipment were kept constant. Each assessment test was performed individually. An instruction session was performed before the tests, in which the volunteer hears two extra speech samples, and the experiment process was explained. Each audio sample received at least 15 scores by the assessors and the scores were averaged to calculate the MOS value. In average, each volunteer scored 93 samples. The speech quality scores obtained on the subjective tests considering two separate groups, male and female speakers, did not present significant average score differences in relation to the score obtained using both gender groups. The coefficient of variation was lower than 1.04\% in each gender group, therefore, an influence of speaker gender was not identified. 
Also, different configuration parameters are used in each test scenario, which are modeled using (9) to obtain \textit{Ppl´}, thus, each parameter has a specific impact on the R-score predictions.

\begin{figure}[!ht]
\centering
\includegraphics[scale=0.44]{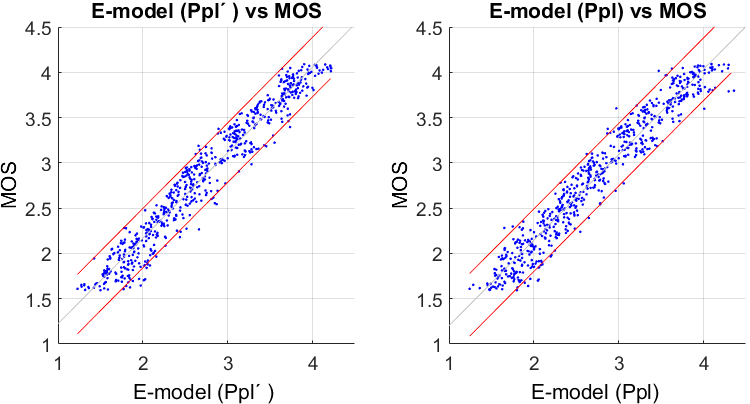}
\caption{Correlation diagrams between MOS values and the results obtained by the E-model (\textit{Ppl´}) and the E-model (\textit{Ppl}).}
\label{scatter}
\vspace{-1mm}
\end{figure}

Based on the subjective test results, the proposed methodology performance was evaluated. As stated before, for each sample, the network scenario configuration was known, from which \textit{Ppl} and \textit{Ppl’} values were determined. In Fig.~\ref{scatter}, the E-model results converted to the 5-point scale are plotted versus the subjective test results (MOS) in a correlation diagram. A confidence interval of 95\% is depicted bounded by red lines. 

As depicted in Fig.~\ref{scatter}, the correlation results between MOS values and the results given by the E-model using \textit{Ppl'} and  \textit{Ppl} are similar, with only few outliers. These similar results are expected because of the high correlation between \textit{Ppl} and \textit{Ppl'} as shown in Table~\ref{tab666}, and Table X presented in the appendix. The PCC and RMSE values between subjective results and E-model predictions are presented in Table~\ref{tab7f}.



\small
\begin{table}[!htbp]
\caption{Performance validation of the proposed methodology using subjective tests.}
\begin{center}
\scalebox{0.84}{%
\resizebox{\columnwidth}{!}{%
\begin{tabular}{c c c }
\hline
 &  \textbf{\textit{PCC}} & \textbf{\textit{RMSE}} \\
\hline
\hline
E-model (\textit{Ppl'}) vs Subjective tests (MOS) & 0.9732& 0.2351  \\ 
E-model (\textit{Ppl}) vs Subjective tests (MOS) & 0.9761& 0.2282  \\ 
\hline
\end{tabular}}%
}
\label{tab7f}
\end{center}
\vspace{-4mm}
\end{table}
\normalsize

As can be observed from Table~\ref{tab7f}, the proposed methodology using the E-model with \textit{Ppl'} reached reliable results; therefore, wireless network parameters could be used for planning purposes that consider speech quality predictions. It is worth mentioning that other wireless network parameters can be included to calculate \textit{Ppl'} and R-scores using the proposed methodology.


\subsection{Final discussions}

Relevant points about the scope and limitations of the proposed methodology are described below:

\begin{itemize}

  \item In principle, the proposed methodology can be applied to the FB E-model, because it is only based on the \textit{Ppl’} used in the equipment impairment factor. However, before extending our methodology to FB networks, the FB E-model needs to be tested in more network scenarios such as those used in the NB and WB E-model versions. It is important to note that currently the FB E-model is only recommended for a specific packet loss model \cite{g1072}. 

    \item In this work, the AWGN channel model was used, represented by the SNR values, but other parameters of different transmission channel models can be used to determine a function to predict \textit{Ppl’} values.
    
    \item In the test scenario, a MIMO system with specific characteristics was implemented and evaluated. However, the proposed methodology is agnostic of the technology used, it only considers the \textit{Ppl’} values. Therefore, different MIMO configurations, $Ms$ or other techniques used in mobile wireless networks can be considered in the proposed methodology. Hence, our proposal can be applied in wireless network planning to estimate speech quality.
    
    \item Experimental tests were performed using only speech signals. Therefore, signals different from speech, such as music, are out of the scope of this work.
    
\end{itemize}

\section{Conclusion}
\label{7}
This work introduced a network simulator that uses different AMR and AMR-WB mode operations, modulation schemes, and channel degradation. With the simulator, it is also possible to determine the actual $Ppl$ values. As first step, the test scenario performance was evaluated using ITU-T Rec. P.863 results as a reference because of the high number of speech samples. Preliminary test results show that in the presence of wireless channel degradation, different antenna arrays of MIMO systems directly impact on the transmission quality in different wireless channel conditions, and this impact can be measured using the number of packet losses. In this context, a novel methodology was proposed to quantitatively determine how the transmission channel and the MIMO system impact on the prediction of R-scores, which is quantified in terms of the proposed $\textit{Ppl'}$ using \eqref{eq7}. Further, the modulation scheme is a relevant factor to determine degradation characteristics, and therefore, the proposed methodology to determine $\textit{Ppl'}$ takes it into consideration. Several scenarios were compared, in which the E-model uses the actual $\textit{Ppl}$ and $\textit{Ppl'}$, and both results in each scenario reached a high correlation. Finally, performance validation results obtained in subjective tests showed that the proposed methodology based on the $\textit{Ppl'}$ parameter of the E-model algorithms permits the estimation of confident R-scores, reaching a PCC and an RMSE of $0.9732$, $0.2351$, respectively. It is noteworthy that other E-model algorithm inputs are not affected. Furthermore, our proposed methodology can be applied to other techniques used in wireless networks. Thus, the E-model algorithms will be useful to predict speech quality in wireless networks.

\appendix 

\section{Results obtained for different Modulation Schemes}
\label{FirstAppendix}

In this appendix, the results of test scenarios that were not treated in the previous sections are presented. In these test scenarios, the BPSK, QAM-16,  QAM-32,  QAM-64,  and  QAM-256 modulation schemes were implemented.

Table~\ref{tapx1} provides the coefficient values of the function used to determine \textit{Ppl'} parameter introduced in (10) using BPSK, QAM-16, QAM-32, QAM-64, and QAM-256 modulation schemes.

\small
\begin{table}[t]
\caption{Coefficient values of the proposed function ($Ppl'= aSNR^b+c$) used to determine \textit{Ppl'} for different $M_s$.}
\begin{center}
\scalebox{0.84}{%
\resizebox{\columnwidth}{!}{%
\begin{tabular}{c c c c c}
\hline
\textbf{\textit{$M_s$}} &  \textbf{\textit{(n$T_x$, m$R_x$)}} & \textbf{\textit{a}} & \textbf{\textit{b}} & \textbf{\textit{c}}\\
\hline
\hline
BPSK & (1,1) & 1019 & -5.002 & -0.002132 \\ 
BPSK  & (2,2) & 90.26 & -4.357 & -0.001042 \\ 
BPSK & (3,3) & 0.1684 & -1.757 & -0.0009269  \\ 
BPSK &  (4,4)  & 0.3692 & -3.25 & -0.000051 \\ \hline
QAM-16 & (1,1) & $1.241\times10^{16}$ & -14.19 & 0 \\ 
QAM-16   & (2,2) & $2.04\times10^{8}$ & -7.805 & -0.0025011 \\ 
QAM-16  & (3,3) & $3.751\times^{14}$ & -14.66 & -0.0000017  \\ 
QAM-16  &  (4,4)  & $4.624\times10^{7}$ & -8.614 & -0.0002708 \\ \hline
QAM-32 & (1,1) & $2.898\times10^{20}$ & -1.78 & -0.0000001 \\ 
QAM-32  & (2,2) & $6.607\times10^{7}$ & -7.132 & -0.009703 \\ 
QAM-32 & (3,3) & $3.725\times10^{10}$ & -9.859 & 0  \\ 
QAM-32 &  (4,4)  & $1.142v10^{15}$ & -14.51 & 0 \\ \hline
QAM-64 & (1,1) & $7.08\times10^{33}$ & -25.91 & 0 \\ 
QAM-64  & (2,2) & $2.756\times10^{30}$ & -24.09 & -0.0000013 \\ 
QAM-64 & (3,3) & $5.403\times10^{27}$ & -23.23 & 0  \\ 
QAM-64 &  (4,4)  & $1.043\times10^{10}$ & -9.116 & -0.002523 \\ \hline
QAM-256 & (1,1) & $5.922\times10^{25}$ & -18.52 & 0 \\ 
QAM-256  & (2,2) & $3.302\times10^{17}$ & -13.05 & 0 \\ 
QAM-256 & (3,3) & $5.465\times10^{22}$ & -17.32 & -0.0000012  \\ 
QAM-256 &  (4,4)  & $2.975\times10^{23}$ & -18.28 & 0 \\ 
\hline
\end{tabular}}%
}
\label{tapx1}
\end{center}
\end{table}
\normalsize

Table~\ref{tapx2} presents the performance validation results of the function defined in (10) using the coefficients presented in Table~\ref{tapx1} for BPSK, QAM-16, QAM-32, QAM-64, and QAM-256 modulation schemes.

\begin{table}[t]
\caption{Performance validation results of the proposed function ($Ppl'= aSNR^b+c$) for different $M_s$.}
\begin{center}
\scalebox{0.64}{%
\resizebox{\columnwidth}{!}{%
\begin{tabular}{c c c c }
\hline
\textbf{\textit{$M_s$}} &  \textbf{\textit{(n$T_x$, m$R_x$)}} & \textbf{\textit{$R^2$}} & \textbf{\textit{RMSE}} \\
\hline
\hline
BPSK & (1,1) & 99.21 & 0.006287  \\ 
BPSK  & (2,2) & 98.75 & 0.007928  \\ 
BPSK & (3,3) & 98.66 & 0.002291  \\ 
BPSK &  (4,4)  & 97.77 & 0.005149  \\ \hline
QAM-16 & (1,1) & 99.55  & 0.002609  \\ 
QAM-16   & (2,2) &  99.38 & 0.005615  \\ 
QAM-16  & (3,3) &  99.64 & 0.002014   \\ 
QAM-16  &  (4,4)  & 99.06 & 0.006421 \\ \hline
QAM-32 & (1,1) & 99.18 & 0.006921  \\ 
QAM-32  & (2,2) & 98.67 &  0.002102 \\ 
QAM-32 & (3,3) & 0.9876 & 0.008115   \\ 
QAM-32 &  (4,4)  & 99.71 & 0.003185 \\ \hline
QAM-64 & (1,1) & 99.78 & 0.003883  \\ 
QAM-64  & (2,2) & 99.32 & 00.8143  \\ 
QAM-64 & (3,3) & 99.92 & 0.001705  \\ 
QAM-64 &  (4,4)  & 99.64 & 0.006176  \\ \hline
QAM-256 & (1,1) & 99.09 & 0.001742  \\ 
QAM-256  & (2,2) & 98.63 & 0.001475  \\ 
QAM-256 & (3,3) & 97.63 & 0.009546  \\ 
QAM-256 &  (4,4)  & 99.73 & 0.003614  \\ \hline
\hline
\end{tabular}}%
}
\label{tapx2}
\end{center}
\end{table}

Table~\ref{tapx3} shows the performance validation results obtained by the appropriate E-model algorithm and using the actual \textit{Ppl} and the estimated \textit{Ppl} (\textit{Ppl’}), considering values corresponding to all antenna set configurations for each speech codec mode operation and using BPSK, QAM-16, QAM-32, QAM-64, and QAM-256 modulation schemes. For NB context, the BurstR parameter introduced in (8) is considered equal to $1$ because the packet loss in the network scenarios is random.

\begin{table}[t]
\caption{Performance validation results of the E-model (\textit{Ppl’}) in relation to the E-model (\textit{Ppl})  considering all the antenna set configurations for each speech codec mode and using different $M_s$.}
\begin{center}
\scalebox{0.70}{%
\resizebox{\columnwidth}{!}{%
\begin{tabular}{c c c c}
\hline
\textbf{\textit{$M_s$}} & \textbf{\textit{Speech codec}} &  \textbf{\textit{PCC}} & \textbf{\textit{RMSE}} \\
\hline
\hline
BPSK & AMR Mode 4 & 0.9855& 2.0021  \\ 
BPSK &AMR Mode 7  & 0.9781 & 2.4450  \\ 
BPSK & AMR-WB Mode 2 & 0.9825 & 2.1866  \\ 
BPSK & AMR-WB Mode 8 &  0.9823 & 2.2684  \\ \hline
QAM-16 & AMR Mode 4 & 0.9732& 2.8792  \\ 
QAM-16 &AMR Mode 7  & 0.9785 & 2.9752  \\ 
QAM-16 &AMR-WB Mode 2 & 0.9810& 2.6217  \\ 
QAM-16 &AMR-WB Mode 8 &  0.9805 & 2.8794  \\ \hline
QAM-32 & AMR Mode 4 & 0.9879& 1.9856  \\ 
QAM-32 &AMR Mode 7  & 0.9782 & 2.5463 \\ 
QAM-32 &AMR-WB Mode 2 & 0.9758 & 3.0516  \\ 
QAM-32 &AMR-WB Mode 8 &  0.9855 & 2.3378  \\  \hline
QAM-64 & AMR Mode 4 & 0.9705& 2.4916 \\ 
QAM-64 &AMR Mode 7  & 0.9785 & 2.8264  \\ 
QAM-64 &AMR-WB Mode 2 & 0.9805 & 2.8875  \\ 
QAM-64 &AMR-WB Mode 8 &  0.9727 & 2.3481  \\ 
\hline
\end{tabular}}%
}
\label{tapx3}
\end{center}
\end{table}

\ifCLASSOPTIONcaptionsoff
  \newpage
\fi

\newpage
\bibliographystyle{IEEEtran}
\bibliography{bibli}

\begin{IEEEbiography}[{\includegraphics[width=1in,height=1.25in,clip,keepaspectratio]{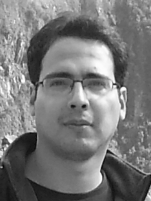}}]{Dem\'ostenes Z. Rodr\'iguez} (M'12-SM'15) received the B.S. degree in electronic engineering from the Pontifical Catholic University of Peru, the M.S. degree and the Ph.D. degree from the University of S\~ao Paulo in 2009 and 2013, respectively. He is currently an Adjunct Professor with the Department of Computer Science, Federal University of Lavras, Brazil. He has a solid knowledge in Telecommunication Systems and Computer Science based on 15 years of professional experience in major companies. His research interests include QoS and QoE in multimedia services and architect solutions in telecommunication systems. He is a member of the Brazilian Telecommunications Society.
\end{IEEEbiography}

\begin{IEEEbiography}[{\includegraphics[width=1in,height=1.25in,clip,keepaspectratio]{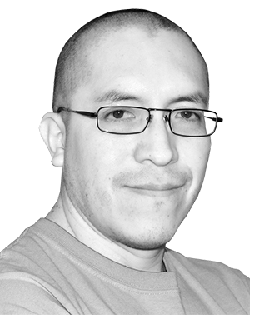}}]{Dick Carrillo Melgarejo} (M'08) received the B.Eng. degree (Hons.) in electronics and electrical engineering from San Marcos National University, Lima, Per\'u, and the M.Sc. degree in electrical engineering from Pontifical Catholic University of Rio de Janeiro, Brazil, in 2004 and 2008, respectively. Between 2008 and 2018 he participated in many projects in the field of research and development of cellular networks, such as LTE, LTE-A, and LTE-A Pro. Since 2018 he is a researcher at LUT University, where he is also pursuing the Ph.D degree in electrical engineering. His research interests are mobile technologies beyond 5G, energy harvesting, intelligent meta-surfaces, and  Cell-free mMIMO. 
\end{IEEEbiography}

\begin{IEEEbiography}[{\includegraphics[width=1in,height=1.25in,clip,keepaspectratio]{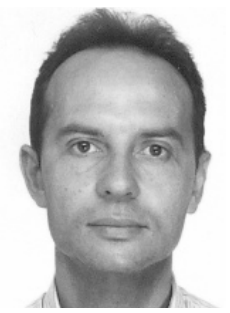}}]{Miguel A. Ramírez} (M'78-SM'00) received the B.S. degree in electronics engineering from the Instituto Tecnológico de Aeronáutica, Brazil, in 1980, the graduate degree in electronic design engineering from the Philips International Institute, The Netherlands, in 1981, and the M.S. and Ph.D. degrees in electrical engineering from the University of São Paulo, Brazil, in 1992 and 1997, respectively. In 2008, he carried out research at the Royal Institute of Technology in Sweden. He is currently an Associate Professor with Escola Politécnica, University of São Paulo. His research focuses on the application of novel signal processing and machine learning algorithms to signal compression and prediction, speech analysis. He is a member of the Brazilian Telecommunications Society.  
\end{IEEEbiography}

\begin{IEEEbiography}
[{\includegraphics[width=1in,height=1.25in,clip,keepaspectratio]{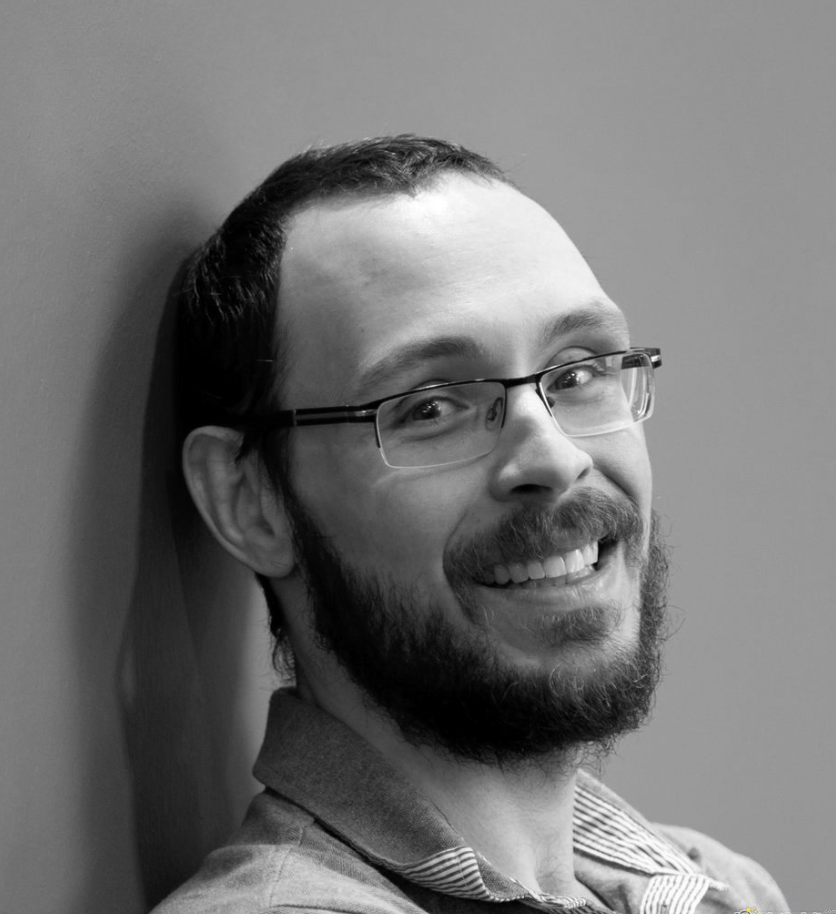}}]
{Pedro H. J. Nardelli} received the B.S. and M.Sc. degrees in electrical engineering from the State University of Campinas, Brazil, in 2006 and 2008, respectively. In 2013 he received his doctoral degree from University of Oulu, Finland, and State University of Campinas following a dual-degree agreement. Nowadays he is assistant professor in IoT in energy systems (tenure track) at the Laboratory of Control Engineering and Digital Systems, School of Energy Systems, Lappeenranta University of Technology, Finland, as well as adjunct professor (docent) in information processing and communication strategies for energy systems at Centre for Wireless Communications, University of Oulu. 
\end{IEEEbiography}

\begin{IEEEbiography}[{\includegraphics[width=1in,height=1.25in,clip,keepaspectratio]{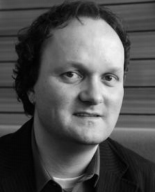}}]{Sebastian M\"oller} studied electrical engineering at the University of Bochum, Bochum, Germany; University of Orleans, Orleans cedex, France; and University of Bologna, Bologna, Italy. He received the Doctor-of-Engineering degree in 1999 and the Venia Legendi with a book on the quality of telephone-based spoken dialog systems in 2004. In 2005, he joined Deutsche Telekom Laboratories, Technical University of Berlin (TU Berlin), Berlin, Germany, and in 2007, he was appointed Professor for Quality and Usability at TU Berlin. His primary interests are in speech signal processing, speech technology, and quality and usability evaluation.
\end{IEEEbiography}

\end{document}

%% file: acronyms.tex
\begin{acronym}
  \acro{1G}{first generation of mobile network}
  \acro{1PPS}{1 pulse per second}
  \acro{2G}{second generation of mobile network}
  \acro{3G}{third generation of mobile network}
  \acro{4G}{fourth generation of mobile network}
  \acro{5G}{fifth generation of mobile network}
  \acro{ARQ}{Automatic repeat request}
  \acro{ASIP}{Application Specific Integrated Processors}
  \acro{AWGN}{additive white Gaussian noise}
   \acro{BER}{bit error rate}
  \acro{BCH}{Bose-Chaudhuri-Hocquenghem}
  \acro{BRIC}{Brazil-Russia-India-China}
  \acro{BS}{base station}
  \acro{CDF}{Cumulative Density Function}
  \acro{CoMP} {cooperative multi-point}
  \acro{CP}{cyclic prefix}
  \acro{CR}{cognitive radio}
  \acro{CS}{cyclic suffix}
  \acro{CSI}{channel state information}
  \acro{CSMA}{carrier sense multiple access}
  \acro{DFT}{discrete Fourier transform}
  \acro{DFT-s-OFDM}{DFT spread OFDM}
  \acro{DSA}{dynamic spectrum access}
  \acro{DVB}{digital video broadcast}
  \acro{DZT}{discrete Zak transform}
  \acro{eMBB} {Enhanced Mobile Broadband}
  \acro{EPC}{evolved packet core}
  \acro{FBMC}{filterbank multicarrier}
  \acro{FDE}{frequency-domain equalization}
  \acro{FDMA}{frequency division multiple access}
  \acro{FD-OQAM-GFDM}{frequency-domain OQAM-GFDM}
  \acro{FEC}{forward error control}
  \acro{F-OFDM}{Filtered Orthogonal Frequency Division Multiplexing}
  \acro{FPGA}{Field Programmable Gate Array}
  \acro{FTN}{Faster than Nyquist}
  \acro{FT}{Fourier transform}
  \acro{FSC}{frequency-selective channel}
  \acro{GFDM}{Generalized Frequency Division Multiplexing}
  \acro{GPS}{global positioning system}
  \acro{GS-GFDM}{guard-symbol GFDM}
  \acro{IARA}{Internet Access for Remote Areas}
  \acro{ICI}{intercarrier interference}
  \acro{IDFT}{Inverse Discrete Fourier Transform}
  \acro{IFI}{inter-frame interference}
  \acro{IMS}{IP multimedia subsystem}
  \acro{IoT}{Internet of Things}
  \acro{IP}{Internet Protocol}
  \acro{ISI}{intersymbol interference}
  \acro{IUI}{inter-user interference}
  \acro{LDPC}{low-density parity check}
  \acro{LLR}{log-likelihood ratio}
  \acro{LMMSE}{linear minimum mean square error}
  \acro{LTE}{long-term evolution}
  \acro{LTE-A}{long-term evolution - advanced}
  \acro{LTE-A Pro}{long-term evolution - advanced pro}
  \acro{M2M}{Machine-to-Machine}
  \acro{MA}{multiple access}
  \acro{MAR}{mobile autonomous reporting}
  \acro{MF}{Matched filter}
  \acro{MIMO}{multiple-input multiple-output}
  \acro{MMSE}{minimum mean square error}
  \acro{MRC}{maximum ratio combiner}
  \acro{MSE}{mean-squared error}
  \acro{MTC}{Machine-Type Communication}
  \acro{NEF}{noise enhancement factor}
  \acro{NFV}{network functions virtualization}
  \acro{OFDM}{Orthogonal Frequency Division Multiplexing}
  \acro{OOB}{out-of-band}
  \acro{OOBE}{out-of-band emission}
  \acro{OQAM}{Offset Quadrature Amplitude Modulation}
  \acro{PAPR}{Peak to average power ratio}
  \acro{PDF}{probability density function}
  \acro{PHY}{physical layer}
  \acro{QAM}{quadrature amplitude modulation}
  \acro{PSD}{power spectrum density}
  \acro{QoE}{quality of experience}
  \acro{QoS}{quality of service}
  \acro{RC}{raised cosine}
  \acro{RRC}{root raised cosine}
  \acro{RTT} {round trip time}  
  \acro{SC}{single carrier}
  \acro{SC-FDE}{Single Carrier Frequency Domain Equalization}
  \acro{SC-FDMA}{Single Carrier Frequency Domain Multiple Access}
  \acro{SDN}{software-defined network}
  \acro{SDR}{software-defined radio}
  \acro{SDW}{software-defined waveform}
  \acro{SEP}{symbol error probability}
  \acro{SER}{symbol error rate}
  \acro{SIC}{successive interference cancellation}
  \acro{SINR}{signal-to-interference-and-noise ratio }
  \acro{SMS}{Short Message Service}
  \acro{SNR}{signal-to-noise ratio}
  \acro{STC}{space time code}
  \acro{STFT}{short-time Fourier transform}
  \acro{TD-OQAM-GFDM}{time-domain OQAM-GFDM}
  \acro{TTI}{time transmission interval}
  \acro{TR-STC}{Time-Reverse Space Time Coding}
  \acro{TR-STC-GFDMA}{TR-STC Generalized Frequency Division Multiple Access}
  \acro{TVC}{ime-variant channel}
  \acro{UFMC}{universal filtered multi-carrier}
  \acro{UF-OFDM}{Universal Filtered Orthogonal Frequency Multiplexing}
  \acro{UHF}{ultra high frequency}
  \acro{URLL}{Ultra Reliable Low Latency}
  \acro{V2V}{vehicle-to-vehicle}
  \acro{V-OFDM}{Vector OFDM}
  \acro{ZF}{zero-forcing}
  \acro{ZMCSC}{zero-mean circular symmetric complex Gaussian}
  \acro{W-GFDM}{windowed GFDM}
  \acro{WHT}{Walsh-Hadamard Transform}
  \acro{WLAN}{wireless Local Area Network}
  \acro{WLE}{widely linear equalizer}
  \acro{WLP}{wide linear processing}
  \acro{WRAN}{Wireless Regional Area Network}
  \acro{WSN}{wireless sensor networks}
  \acro{ROI}{return on investment}
  \acro{NR}{new radio}
  \acro{SAE}{system architecture evolution}
  \acro{E-UTRAN}{evolved UTRAN}
  \acro{3GPP}{3rd Generation Partnership Project }
  \acro{MME}{mobility management entity}
  \acro{S-GW}{serving gateway}
  \acro{P-GW}{packet-data network gateway}
  \acro{eNodeB}{evolved NodeB}
  \acro{UE}{user equipment}
  \acro{DL}{downlink}
  \acro{UL}{uplink}
  \acro{LSM}{link-to-system mapping}
  \acro{PDSCH}{physical downlink shared channel}
  \acro{TB}{transport block}
  \acro{MCS}{modulation code scheme}
  \acro{ECR}{effective code rate}
  \acro{BLER}{block error rate}
  \acro{CCI}{co-channel interference}
  \acro{OFDMA}{orthogonal frequency-division multiple access}
  \acro{LOS}{line-of-sight}
  \acro{VHF}{very high frequency}
  \acro{pdf}{probability density function}
  \acro{ns-3}{Network simulator 3}
  \acro{Mbps}{mega bits per second}
  \acro{IPC}{industrial personal computer}
  \acro{RSSI}{received signal strength indicator}
  \acro{OPEX}{operational expenditures}
  \acro{H2H}{human-to-human}
  \acro{DOD}{depth of discharge}
  \acro{ADC}{analog-to-digital converter}
  \acro{FFT}{Fourier fast transform}
  \acro{DAC}{digital-to-analog converter}
  \acro{IFFT}{inverse Fourier fast transform}
  \acro{DC/DC}{Direct-to-Direct current converter}
  \acro{BBI}{brain-to-brain interface}
  \acro{BBC}{brain-to-brain communication}
  \acro{SON}{self-organized networks}
  \acro{NOMA}{non-orthogonal multiple access}
  \acro{UMTS}{universal mobile telecommunications system}
  \acro{HSDPA}{high-speed downlink packet access}
  \acro{OSTBC}{orthogonal space-time block code}

\end{acronym}

%% file: main.bbl
\begin{thebibliography}{10}
\providecommand{\url}[1]{#1}
\csname url@samestyle\endcsname
\providecommand{\newblock}{\relax}
\providecommand{\bibinfo}[2]{#2}
\providecommand{\BIBentrySTDinterwordspacing}{\spaceskip=0pt\relax}
\providecommand{\BIBentryALTinterwordstretchfactor}{4}
\providecommand{\BIBentryALTinterwordspacing}{\spaceskip=\fontdimen2\font plus
\BIBentryALTinterwordstretchfactor\fontdimen3\font minus
  \fontdimen4\font\relax}
\providecommand{\BIBforeignlanguage}[2]{{%
\expandafter\ifx\csname l@#1\endcsname\relax
\typeout{** WARNING: IEEEtran.bst: No hyphenation pattern has been}%
\typeout{** loaded for the language `#1'. Using the pattern for}%
\typeout{** the default language instead.}%
\else
\language=\csname l@#1\endcsname
\fi
#2}}
\providecommand{\BIBdecl}{\relax}
\BIBdecl

\bibitem{Montag2015}
C.~Montag, K.~B{\l}aszkiewicz, R.~Sariyska, B.~Lachmann, I.~Andone,
  B.~Trendafilov, M.~Eibes, and A.~Markowetz, ``Smartphone usage in the 21st
  century: who is active on whatsapp?'' \emph{BMC Research Notes}, vol.~8,
  no.~1, pp. 331--336, Aug. 2015.

\bibitem{CiscoForecast}
{Cisco Inc.}, ``Visual networking index: Global mobile data traffic forecast
  update, 2016–2021,'' Jun. 2017.

\bibitem{raake_terminal}
M.~{Soloducha}, A.~{Raake}, F.~{Kettler}, and S.~{Bleiholder}, ``Testing
  conversational quality of {VoIP} with different terminals and degradations,''
  in \emph{Ninth International Conference on Quality of Multimedia Experience
  (QoMEX)}, May 2017, pp. 1--3.

\bibitem{raake_intell}
F.~{Schiffner}, J.~{Skowronek}, and A.~{Raake}, ``On the impact of speech
  intelligibility on speech quality in the context of voice over {IP}
  telephony,'' in \emph{Sixth International Workshop on Quality of Multimedia
  Experience (QoMEX)}, Sep. 2014, pp. 59--60.

\bibitem{itup800}
\BIBentryALTinterwordspacing
{ITU-T Rec. P.800}, ``Methods for subjective determination of transmission
  quality,'' Jun. 1996. [Online]. Available:
  \url{http://www.itu.int/rec/T-REC-P.800/en.}
\BIBentrySTDinterwordspacing

\bibitem{sebastian}
S.~M\"oller, W.~Y. Chan, N.~C\^ot\'e, T.~H. Falk, A.~Raake, and
  M.~W\"altermann, ``Speech quality estimation: Models and trends,'' \emph{IEEE
  Signal Processing Magazine}, vol.~28, no.~6, pp. 18--28, Nov 2011.

\bibitem{itu}
\BIBentryALTinterwordspacing
{ITU-T Rec. P.862}, ``Perceptual evaluation of speech quality ({PESQ}): An
  objective method for end-to-end speech quality assessment of narrow-band
  telephone networks and speech codecs,'' Feb. 2001. [Online]. Available:
  \url{http://www.itu.int/rec/T-REC-P.862/en}
\BIBentrySTDinterwordspacing

\bibitem{itup8622}
\BIBentryALTinterwordspacing
{ITU-T Rec. P.862.2}, ``Wideband extension to recommendation {P.862} for the
  assessment of wideband telephone networks and speech codecs,'' Nov. 2007.
  [Online]. Available: \url{https://www.itu.int/rec/T-REC-P.862.2-200711-I/en}
\BIBentrySTDinterwordspacing

\bibitem{ITUTP863}
\BIBentryALTinterwordspacing
{ITU-T Rec. P.863}, ``Perceptual objective listening quality assessment
  ({POLQA}),'' Aug. 2018. [Online]. Available:
  \url{https://www.itu.int/rec/T-REC-P.863-201803-I/en}
\BIBentrySTDinterwordspacing

\bibitem{beerends2013}
J.~G. Beerends, C.~Schmidmer, J.~Berger, M.~Obermann, R.~Ullmann, J.~Pomy, and
  M.~Keyhl, ``Perceptual objective listening quality assessment ({POLQA}), the
  third generation itu-t standard for end-to-end speech quality measurement
  part i—temporal alignment,'' \emph{J. Audio Eng. Soc}, vol.~61, no.~6, pp.
  366--384, 2013.

\bibitem{polakyy}
J.~Polacky and P.~Pocta, ``An analysis of the impact of packet loss, codecs and
  type of voice on internal parameters of {P.563} model,'' in \emph{Proc. Int.
  Conf. on Digital Technlogies}, Zilina, Slovakia, Jul. 2014, pp. 281--284.

\bibitem{abareg}
M.~Abareghi, M.~Homayounpour, M.~Dehghan, and A.~Davoodi, ``Improved itu-p.563
  non-intrusive speech quality assessment method for covering {VoIP}
  conditions,'' in \emph{Proc. IEEE Int. Conf. on Advanced Communication
  Technology}, South Korea, Feb. 2008, pp. 281--284.

\bibitem{pocta_plr}
J.~{Polacky}, P.~{Pocta}, and R.~{Jarina}, ``Influence of packet loss on a
  speaker verification system over ip network,'' in \emph{International
  Conference Radioelektronika (RADIOELEKTRONIKA)}, April 2016, pp. 390--394.

\bibitem{barriac_access}
M.~A. {Raja}, A.~{Jagodzinska}, and V.~{Barriac}, ``{On Losses, Pauses, Jumps,
  and the Wideband E-Model},'' \emph{IEEE Access}, vol.~5, pp.
  16\,130--16\,148, 2017.

\bibitem{raake_rplr}
M.~{Soloducha} and A.~{Raake}, ``Speech quality of {VoIP}: bursty packet loss
  revisited,'' in \emph{Speech Communication; 11. ITG Symposium}, Sep. 2014,
  pp. 1--4.

\bibitem{mittag_plr}
G.~{Mittag} and S.~{Möller}, ``Single-ended packet loss rate estimation of
  transmitted speech signals,'' in \emph{IEEE Global Conference on Signal and
  Information Processing (GlobalSIP)}, Nov 2018, pp. 226--230.

\bibitem{7776170}
F.~Koester, V.~Cercos-Llombart, G.~Mittag, and S.~Moeller, ``Non-intrusive
  estimation model for the speech-quality dimension loudness,'' in \emph{Proc.
  Conference on Speech Communication}, Oct. 2016, pp. 175--179.

\bibitem{pqs2016}
T.~P. K.~Koster, G.~Mittag and S.~Moeller, ``Non-intrusive estimation of
  noisiness as a perceptual quality dimension of transmitted speech,'' in
  \emph{Proc. Workshop on Perceptual Quality of Systems}, Berlin, Germany, Aug.
  2016, pp. 29--31.

\bibitem{PLR2}
E.~T. Affonso, R.~L. Rosa, and D.~Z. Rodr\'iguez, ``Speech quality assessment
  over lossy transmission channels using deep belief networks,'' \emph{IEEE
  Signal Processing Letters}, vol.~25, no.~1, pp. 70--74, Jan 2018.

\bibitem{visqol_1}
A.~{Hines}, P.~{Počta}, and H.~{Melvin}, ``Detailed comparative analysis of
  {PESQ} and {VISQOL} behaviour in the context of playout delay adjustments
  introduced by {VoIP} jitter buffer algorithms,'' in \emph{Fifth International
  Workshop on Quality of Multimedia Experience}, July 2013, pp. 18--23.

\bibitem{visqol}
C.~Sloan, N.~Harte, D.~Kelly, A.~C. Kokaram, and A.~Hines, ``Objective
  assessment of perceptual audio quality using {ViSQOLAudio},'' \emph{IEEE
  Transactions on Broadcasting}, vol.~63, no.~4, pp. 693--705, Dec 2017.

\bibitem{8468124}
J.~M. Martín-Doñas, A.~M. Gomez, J.~A. Gonzalez, and A.~M. Peinado, ``A deep
  learning loss function based on the perceptual evaluation of the speech
  quality,'' \emph{IEEE Signal Processing Letters}, vol.~25, no.~11, pp.
  1680--1684, Nov 2018.

\bibitem{ietmeu}
R.~{Dantas Nunes}, R.~{Lopes Rosa}, and D.~{Zegarra Rodríguez}, ``Performance
  improvement of a non-intrusive voice quality metric in lossy networks,''
  \emph{IET Communications}, vol.~13, no.~20, pp. 3401--3408, 2019.

\bibitem{8462042}
R.~Fakoor, X.~He, I.~Tashev, and S.~Zarar, ``Constrained
  convolutional-recurrent networks to improve speech quality with low impact on
  recognition accuracy,'' in \emph{IEEE International Conference on Acoustics,
  Speech and Signal Processing (ICASSP)}, April 2018, pp. 3011--3015.

\bibitem{8463408}
M.~Narbutt, A.~Allen, J.~Skoglund, M.~Chinen, and A.~Hines, ``{AMBIQUAL} - a
  full reference objective quality metric for ambisonic spatial audio,'' in
  \emph{Proc. International Conference on Quality of Multimedia Experience
  (QoMEX)}, May 2018, pp. 1--6.

\bibitem{8513822}
E.~T. {Affonso}, R.~D. {Nunes}, R.~L. {Rosa}, G.~F. {Pivaro}, and D.~Z.
  {Rodríguez}, ``Speech quality assessment in wireless {VoIP} communication
  using {Deep Belief Network},'' \emph{IEEE Access}, vol.~6, pp.
  77\,022--77\,032, 2018.

\bibitem{Emodel22}
\BIBentryALTinterwordspacing
{ITU-T Rec. G.107}, ``The {E-model}: a computational model for use in
  transmission planning,'' Jun. 2014. [Online]. Available:
  \url{https://www.itu.int/rec/T-REC-G.107.}
\BIBentrySTDinterwordspacing

\bibitem{amrwb}
{3GPP TS 26.171}, ``{Adaptive Multi-Rate - Wideband (AMR-WB) speech codec;
  General Description (v15.0.0)},'' Jun. 2018.

\bibitem{laura}
L.~F. Gallardo, ``Effects of transmitted speech bandwidth on subjective
  assessments of speaker characteristics,'' in \emph{Proc. International
  Conference on Quality of Multimedia Experience (QoMEX)}, Sardinia, Italy, May
  2018, pp. 1--5.

\bibitem{Moller}
S.~M\"oller, N.~Cote, V.~Gautier-Turbin, N.~Kitawaki, and A.~Takahashi,
  ``Instrumental estimation of {E}-model parameters for wideband speech
  codecs,'' \emph{EURASIP Journal on Audio, Speech, and Music Processing},
  vol.~14, no.~6, pp. 1--16, Nov 2010.

\bibitem{g1071}
\BIBentryALTinterwordspacing
{ITU-T Rec. G.107.1}, ``Wideband {E-model},'' Jun. 2015. [Online]. Available:
  \url{https://www.itu.int/rec/T-REC-G.107.1/en.}
\BIBentrySTDinterwordspacing

\bibitem{ieeeaccess}
J.~Abichandani, J.~Baenke, M.~S. Irizarry, N.~Saxena, P.~Vyas, S.~Prasad,
  S.~Mada, and Y.~Z. Tafesse, ``A comparative study of voice quality and
  coverage for voice over long term evolution calls using different codec
  mode-sets,'' \emph{IEEE ACCESS}, vol.~5, no.~6, pp. 10\,315--10\,322, Jun
  2017.

\bibitem{mittag}
G.~Mittag, S.~M\"oller, V.~Barriac, and S.~Ragot, ``Quantifying quality
  degradation of the {EVS Super-Wideband Speech Codec},'' in \emph{Proc.
  International Conference on Quality of Multimedia Experience {(QoMEX)}},
  Sardinia, Italy, May. 2018, pp. 1--6.

\bibitem{barriac_swb}
S.~{Tiemounou}, R.~{Le Bouquin Jeannès}, and V.~{Barriac}, ``Performance
  evaluation of quality degradation indicators on super-wideband speech
  signals,'' in \emph{2012 Proceedings of the 20th European Signal Processing
  Conference (EUSIPCO)}, Aug 2012, pp. 2792--2796.

\bibitem{g1072}
\BIBentryALTinterwordspacing
{ITU-T Rec. G.107.2}, ``Fullband {E-model},'' Jun. 2019. [Online]. Available:
  \url{https://www.itu.int/rec/T-REC-G.107.2/en.}
\BIBentrySTDinterwordspacing

\bibitem{evs}
S.~{Bruhn}, H.~{Pobloth}, M.~{Schnell}, B.~{Grill}, J.~{Gibbs}, L.~{Miao},
  K.~{Järvinen}, L.~{Laaksonen}, N.~{Harada}, N.~{Naka}, S.~{Ragot},
  S.~{Proust}, T.~{Sanda}, I.~{Varga}, C.~{Greer}, M.~{Jelínek}, M.~{Xie}, and
  P.~{Usai}, ``Standardization of the new {3GPP EVS codec},'' in \emph{IEEE
  Int. Conf. on Acoustics, Speech and Signal Processing (ICASSP)}, April 2015,
  pp. 5703--5707.

\bibitem{opus}
J.-M. Valin, K.~Vos, and T.~B. Terriberry, ``Definition of the {Opus Audio
  Codec},'' in \emph{{IETF RFC6716}}, April 2012, pp. 1--326.

\bibitem{pocta_opus}
M.~{Al-Ahmadi}, P.~{Pocta}, and H.~{Melvin}, ``Instrumental estimation of
  {E}-model equipment impairment factor parameters for super-wideband {O}pus
  codec,'' in \emph{Irish Signals and Systems Conf.}, June 2019, pp. 1--5.

\bibitem{R5}
M.~Nicolaou, A.~Doufexi, S.~Armour, and Y.~Sun, ``{Scheduling Techniques for
  Improving Call Capacity for VoIP Traffic in MIMO-OFDMA Networks},'' in
  \emph{{Proc. IEEE 70th Vehicular Technology Conference}}, Anchorage, US,
  Sept. 2009, pp. 1--5.

\bibitem{mimo1}
M.~Yang, L.~Sun, X.~Yuan, and B.~Chen, ``{A New Nested MIMO Array With
  Increased Degrees of Freedom and Hole-Free Difference Coarray},'' \emph{IEEE
  Signal Processing Letters}, vol.~25, no.~1, pp. 40--44, Jan 2018.

\bibitem{mimo2}
A.~Mayouche, A.~Metref, and J.~Choi, ``{Downlink Training Overhead Reduction
  Technique for FDD Massive MIMO Systems},'' \emph{IEEE Signal Processing
  Letters}, vol.~25, no.~8, pp. 1201--1205, Aug 2018.

\bibitem{mimo3}
Y.~Teng, M.~Wei, A.~Liu, V.~Lau, and Y.~Zhang, ``{Mixed-Timescale Per-Group
  Hybrid Precoding for Multiuser Massive MIMO Systems},'' \emph{IEEE Signal
  Processing Letters}, vol.~25, no.~5, pp. 675--679, May 2018.

\bibitem{g108}
\BIBentryALTinterwordspacing
{ITU-T Rec. G.108}, ``{Application of the E-model: A planning guide},'' Sep.
  1999. [Online]. Available: \url{https://www.itu.int/rec/T-REC-G.108.}
\BIBentrySTDinterwordspacing

\bibitem{g1081}
\BIBentryALTinterwordspacing
{ITU-T Rec. G.108.1}, ``{Guidance for assessing conversational speech
  transmission quality effects not covered by the E-model},'' May 2000.
  [Online]. Available: \url{https://www.itu.int/rec/T-REC-G.107.1/en.}
\BIBentrySTDinterwordspacing

\bibitem{g1082}
\BIBentryALTinterwordspacing
{ITU-T Rec. G.108.2}, ``Transmission planning aspects of echo cancellers,''
  Mar. 2007. [Online]. Available:
  \url{https://www.itu.int/rec/T-REC-G.107.1/en.}
\BIBentrySTDinterwordspacing

\bibitem{itug113}
\BIBentryALTinterwordspacing
{ITU-T Rec. G.113}, ``Transmission impairments due to speech processing,'' Mar.
  2007. [Online]. Available:
  \url{https://www.itu.int/rec/T-REC-G.113-200711-I/en}
\BIBentrySTDinterwordspacing

\bibitem{itug113anex04}
\BIBentryALTinterwordspacing
{ITU-T Rec. G.113 - Appendix IV}, ``Revised {Appendix IV} - provisional
  planning values for the wideband equipment impairment factor and the wideband
  packet loss robustness factor,'' Mar. 2009. [Online]. Available:
  \url{https://www.itu.int/rec/T-REC-G.113-200903-I!Amd1/en}
\BIBentrySTDinterwordspacing

\bibitem{itug113anex05}
\BIBentryALTinterwordspacing
{ITU-T Rec. G.113 - Appendix V}, ``New {Appendix V} – provisional planning
  values for the fullband equipment impairment factor and the fullband packet
  loss robustness factor,'' May 2019. [Online]. Available:
  \url{https://www.itu.int/rec/T-REC-G.113-200903-I!Amd1/en}
\BIBentrySTDinterwordspacing

\bibitem{itup834}
\BIBentryALTinterwordspacing
{ITU-T Rec. P.834}, ``Methodology for the derivation of equipment impairment
  factors from instrumental models,'' Jun. 2015. [Online]. Available:
  \url{https://www.itu.int/rec/T-REC-P.833-200102-I/en}
\BIBentrySTDinterwordspacing

\bibitem{p8341}
\BIBentryALTinterwordspacing
{ITU-T Rec. P.834.1}, ``Extension of the methodology for the derivation of
  equipment impairment factors from instrumental models for wideband speech
  codecs,'' Jun. 2015. [Online]. Available:
  \url{https://www.itu.int/rec/T-REC-P.834.1/en.}
\BIBentrySTDinterwordspacing

\bibitem{itup833}
\BIBentryALTinterwordspacing
{ITU-T Rec. P.833}, ``Methodology for derivation of equipment impairment
  factors from subjective listening-only tests,'' Feb. 2001. [Online].
  Available: \url{https://www.itu.int/rec/T-REC-P.833-200102-I/en}
\BIBentrySTDinterwordspacing

\bibitem{itup8331}
\BIBentryALTinterwordspacing
{ITU-T Rec. P.833.1}, ``Methodology for the derivation of equipment impairment
  factors from subjective listening-only tests for wideband speech codecs,''
  Apr. 2009. [Online]. Available:
  \url{https://www.itu.int/rec/T-REC-P.833-200102-I/en}
\BIBentrySTDinterwordspacing

\bibitem{amr}
{3GPP TS 26.071}, ``{AMR Speech Codec; General Description (v15.0.0)},'' Jun.
  2018.

\bibitem{amr22}
{3GPP TS 26.101}, ``{Adaptive Multi-Rate (AMR) Speech Codec - Frame structure
  (v15.0.0)},'' Jun. 2018.

\bibitem{g722_2}
\BIBentryALTinterwordspacing
{ITU-T Rec. G.722.2}, ``{Wideband coding of speech at around 16 kbit/s using
  Adaptive Multi-Rate Wideband (AMR-WB) },'' Jul 2013. [Online]. Available:
  \url{https://www.itu.int/rec/T-REC-G.722.2-200307-I/en}
\BIBentrySTDinterwordspacing

\bibitem{amrperform}
{3GPP TS 26.976}, ``{Performance characterization of the Adaptive Multi-Rate
  Wideband (AMR-WB) speech codec (v15.0.0)},'' Jun. 2018.

\bibitem{siswire}
S.~Haykin and M.~Moher, \emph{Modern wireless communications}.\hskip 1em plus
  0.5em minus 0.4em\relax Upper Saddle River, NJ.: Pearson Prentice Hall, 2005.

\bibitem{6517180}
H.~Lee, S.~Byeon, B.~Kim, K.~B. Lee, and S.~Choi, ``{Enhancing Voice over WLAN
  via Rate Adaptation and Retry Scheduling},'' \emph{IEEE Trans. on Mobile
  Computing}, vol.~13, no.~12, pp. 2791--2805, Dec 2014.

\bibitem{5301913}
K.~S. Shanmugan, ``Simulation-based estimate of {QoS} for voice traffic over
  {WCDMA} radio links,'' in \emph{Proc. International Conference on Wireless
  Communications, Networking and Mobile Computing}, Beijing, China, Sep. 2009,
  pp. 1--4.

\bibitem{6804500}
A.~Khare, K.~Trivedi, and S.~Dixit, ``Effect of doppler frequency and {BER} in
  {FFT} based {OFDM} system with {Rayleigh} fading channel,'' in \emph{Proc.
  {IEEE} Students Conference on Electrical, Electronics and Computer Science
  ({SCEECS}'2014)}, Bhopal, India, Mar. 2014, pp. 1--6.

\bibitem{gibson2002communications}
D.~J. Gibson, \emph{The communications handbook}.\hskip 1em plus 0.5em minus
  0.4em\relax CRC press, 2002.

\bibitem{rician1}
K.~{Jiang}, Z.~{Liu}, P.~{Yang}, Y.~{Xiao}, and S.~{Li}, ``An experimental
  investigation of enhanced sm-ofdm over indoor rician multipath channels,''
  \emph{IEEE Trans. Veh. Technol.}, vol.~69, no.~2, pp. 2291--2295, 2020.

\bibitem{rayle1}
J.~{Qiu}, K.~{Xu}, X.~{Xia}, Z.~{Shen}, and W.~{Xie}, ``Downlink power
  optimization for cell-free massive mimo over spatially correlated {R}ayleigh
  fading channels,'' \emph{IEEE Access}, vol.~8, pp. 56\,214--56\,227, 2020.

\bibitem{745083422}
M.~Abo-Zahhad, M.~Farrag, and A.~Ali, ``A fast accurate method for calculating
  symbol error probabilities for {AWGN} and {R}ayleigh fading channels,'' in
  \emph{Proc. National Radio Science Conference (NRSC)}, Feb 2016, pp.
  241--248.

\bibitem{awgn222}
A.~{El Gamal}, M.~{Mohseni}, and S.~{Zahedi}, ``Bounds on capacity and minimum
  energy-per-bit for {AWGN} relay channels,'' \emph{IEEE Transactions on
  Information Theory}, vol.~52, no.~4, pp. 1545--1561, 2006.

\bibitem{6770094}
G.~J. {Foschini}, ``Layered space-time architecture for wireless communication
  in a fading environment when using multi-element antennas,'' \emph{Bell Labs
  Technical Journal}, vol.~1, no.~2, pp. 41--59, 1996.

\bibitem{825818}
B.~M. {Hochwald} and T.~L. {Marzetta}, ``Unitary space-time modulation for
  multiple-antenna communications in {R}ayleigh flat fading,'' \emph{IEEE
  Transactions on Information Theory}, vol.~46, no.~2, pp. 543--564, 2000.

\bibitem{4339800}
J.~{Wang}, Y.~{Zhao}, and S.~{Fan}, ``Unitary space-time modulation of turbo
  trellis-coded for multiple-antenna {R}ayleigh fading channel,'' in \emph{2007
  International Conference on Wireless Communications, Networking and Mobile
  Computing}, 2007, pp. 72--76.

\bibitem{730453}
S.~M. {Alamouti}, ``A simple transmit diversity technique for wireless
  communications,'' \emph{IEEE Journal on Selected Areas in Communications},
  vol.~16, no.~8, pp. 1451--1458, 1998.

\bibitem{1374943}
{Aijun Song}, {Genyuan Wang}, {Weifeng Su}, and {Xiang-Gen Xia}, ``Unitary
  space-time codes from alamouti's scheme with apsk signals,'' \emph{IEEE
  Trans. on Wireless Communications}, vol.~3, no.~6, pp. 2374--2384, 2004.

\bibitem{proakiss}
J.~Proakis and M.~Salehi, \emph{Digital Communications}.\hskip 1em plus 0.5em
  minus 0.4em\relax New York, NY, USA: Mc Graw-Hill, 2008.

\bibitem{Tse}
D.~Tse and P.~Viswanath, \emph{Fundamentals of Wireless Communication}.\hskip
  1em plus 0.5em minus 0.4em\relax New York, NY, US: Cambridge University
  Press, 2005.

\bibitem{itup501}
\BIBentryALTinterwordspacing
{ITU-T Rec. P.501}, ``Test signals for use in telephonometry,'' Mar. 2017.
  [Online]. Available: \url{https://www.itu.int/rec/T-REC-P.501-201703-I/en}
\BIBentrySTDinterwordspacing

\bibitem{itup8631}
\BIBentryALTinterwordspacing
{ITU-T Rec. P.863.1}, ``Application guide for recommendation {ITU-T P.863},''
  Sep. 2014. [Online]. Available:
  \url{https://www.itu.int/rec/T-REC-P.863.1/en.}
\BIBentrySTDinterwordspacing

\bibitem{itup341}
\BIBentryALTinterwordspacing
{ITU-T Rec. P.341}, ``Transmission characteristics for wideband digital
  loudspeaking and hands-free telephony terminals,'' Mar. 2012. [Online].
  Available: \url{https://www.itu.int/rec/T-REC-P.341-201103-I/en.}
\BIBentrySTDinterwordspacing

\bibitem{itup191}
\BIBentryALTinterwordspacing
{ITU-T Rec. P.191}, ``Software tools for speech and audio coding
  standardization,'' Mar. 2010. [Online]. Available:
  \url{https://www.itu.int/rec/T-REC-G.191-201003-I/en.}
\BIBentrySTDinterwordspacing

\bibitem{amr2}
{3GPP TS 26.201}, ``{Adaptive Multi-Rate - Wideband (AMR-WB) speech codec -
  Frame structure},'' Jun. 2018.

\bibitem{ostbc}
J.~G. Proakis and M.~Salehi, \emph{Digital Communications}.\hskip 1em plus
  0.5em minus 0.4em\relax New York, NY, USA: Mc Graw-Hill, 2008.

\bibitem{5759942}
F.~{Mertz} and P.~{Vary}, ``Efficient voice communication inwireless packet
  networks,'' in \emph{ITG Conference on Voice Communication [8.
  ITG-Fachtagung]}, 2008, pp. 1--4.

\end{thebibliography}
